\documentclass[journal=jctc,manuscript=article]{achemso}

\usepackage{chemformula} 
\usepackage[T1]{fontenc} 
\usepackage{physics}
\usepackage{tikz}
\usetikzlibrary{decorations.pathreplacing,calligraphy}
\usetikzlibrary{arrows.meta}
\usepackage{array}
\usetikzlibrary{fit}

\newcommand*{\citen}[1]{%
  \begingroup
    \romannumeral-`\x 
    \setcitestyle{numbers}%
    \cite{#1}%
  \endgroup
}
\SectionNumbersOn
\author{Dilhan Manawadu}
\affiliation[]
{Department of Chemistry, Physical and Theoretical Chemistry Laboratory, \\University of Oxford, Oxford, OX1 3QZ, United Kingdom}
\alsoaffiliation[Second University]
{Linacre College, University of Oxford, Oxford, OX1 3JA, United Kingdom}
\email{dilhan.manawadu@chem.ox.ac.uk}
\author{Darren J. Valentine}
\affiliation[]
{Department of Chemistry, Physical and Theoretical Chemistry Laboratory, \\University of Oxford, Oxford, OX1 3QZ, United Kingdom}
\alsoaffiliation[Second University]
{Balliol College, University of Oxford, Oxford, OX1 3BJ, United Kingdom}

\author{William Barford}
\affiliation[]
{Department of Chemistry, Physical and Theoretical Chemistry Laboratory, \\University of Oxford, Oxford, OX1 3QZ, United Kingdom}
\email{william.barford@chem.ox.ac.uk}

\graphicspath{{./figures/}}
\title[An \textsf{achemso} demo]
  {Dynamical simulations of carotenoid photoexcited states using density matrix renormalization group techniques}

\abbreviations{IR,NMR,UV}

\begin{document}

\begin{tocentry}

\resizebox{8.4cm}{4.5cm}{%
\begin{tikzpicture}[
    font=\sffamily,
    level/.style={black,thick},
    sublevel/.style={black,densely dashed},
    ionization/.style={black,dashed},
    transition/.style={black,->,>=stealth',shorten >=1pt},
    radiative/.style={transition,decorate,decoration={snake,amplitude=1.5}},
    indirectradiative/.style={radiative,densely dashed},
    nonradiative/.style={transition,dashed},
    thick,scale=0.6, every node/.style={scale=0.6},
  ]

%
%
%

  \tikzstyle{every node}=[font=\small]
\coordinate (A1) at (-10,10);
\coordinate (A2) at (-9,9);
\coordinate (B) at (-8,9.5);
\coordinate (C) at (-6,9.5);
\coordinate (D1) at (-5,10);
\coordinate (D2) at (8,9);

  \draw[fill=red] (A1) rectangle (A2);
  \draw (B) circle (0.5cm);
  \draw (C) circle (0.5cm);
  \draw[fill=green] (D1) rectangle (D2);

    \node[align=center,font=\small] at (-9.5,9.5) {$1$};
\node[align=center,font=\small] at (1.5,9.5) {$N-3$};

    \coordinate (P1) at (-9.5,10.2);
   \coordinate (P2) at  (-8,10.2);

  \draw [decorate,
    decoration = {calligraphic brace}] (P1) --  +(1.5,0);
  \draw [decorate,
    decoration = {calligraphic brace}] (P2) --  +(2,0);

  \node[align=center] at (-8.75,11) {$\hat{U}_{1} $};
   \node[align=center] at (-7,11) {$\hat{U}_{2} $};


   \coordinate (A1) at (-10,7);
\coordinate (A2) at (-7,6);
\coordinate (B) at (-6,6.5);
\coordinate (C) at (-4,6.5);
\coordinate (D1) at (-3,7);
\coordinate (D2) at (8,6);

  \draw[fill=red] (A1) rectangle (A2);
  \draw (B) circle (0.5cm);
  \draw (C) circle (0.5cm);
  \draw[fill=green](D1) rectangle (D2);

   \node[align=center,font=\small] at (-8.5,6.5) {$2$};
\node[align=center,font=\small] at (2.5,6.5) {$N-4$};

    \coordinate (P1) at (-7.5,7.2);
   \coordinate (P2) at  (-6,7.2);

  \draw [decorate,
    decoration = {calligraphic brace}] (P2) --  +(2,0);

   \node[align=center] at (-5,8) {$\hat{U}_{3}$};

      \draw [-{Stealth}{Stealth}{Stealth}] (-1,5.9)--(-1,4.5);


  \coordinate (A1) at (-10,3);
\coordinate (A2) at (-3,2);
\coordinate (B) at (-2,2.5);
\coordinate (C) at (0,2.5);
\coordinate (D1) at (1,3);
\coordinate (D2) at (8,2);

  \draw[fill=red] (A1) rectangle (A2);
  \draw (B) circle (0.5cm);
  \draw (C) circle (0.5cm);
  \draw[fill=green] (D1) rectangle (D2);

    \coordinate (P1) at (-3.5,3.2);
   \coordinate (P2) at  (-2,3.2);

  \draw [decorate,
    decoration = {calligraphic brace}] (P2) --  +(2,0);

   \node[align=center] at (-1,4) {$\hat{U}_{n} $};

      \node[align=center,font=\small] at (-6.5,2.5) {$n-1$};
\node[align=center,font=\small] at (4.5,2.5) {$N-n-1$};


  \coordinate (A1) at (-10,0);
\coordinate (A2) at (3,-1);
\coordinate (B) at (4,-0.5);
\coordinate (C) at (6,-0.5);
\coordinate (D1) at (7,0);
\coordinate (D2) at (8,-1);

  \draw[fill=red] (A1) rectangle (A2);
  \draw (B) circle (0.5cm);
  \draw (C) circle (0.5cm);
  \draw[fill=green] (D1) rectangle (D2);

    \coordinate (P1) at (4,0.2);
   \coordinate (P2) at  (6,0.2);

  \draw [decorate,
    decoration = {calligraphic brace}] (P1) --  +(2,0);
  \draw [decorate,
    decoration = {calligraphic brace}] (P2) --  +(1.5,0);

  \node[align=center] at (5,1) {$\hat{U}_{N-2} $};
   \node[align=center] at (6.75,1) {$\hat{U}_{N-1} $};

         \node[align=center,font=\small] at (-3.5,-0.5) {$N-3$};
\node[align=center,font=\small] at (7.5,-0.5) {$1$};

\end{tikzpicture}
}

%
%
%

\end{tocentry}

\begin{abstract}
We  present a dynamical simulation scheme to model the highly correlated excited state dynamics of linear polyenes. We apply it to investigate the internal conversion processes of carotenoids following  their photoexcitation. We use the extended Hubbard-Peierls model, $\hat{H}_{\textrm{UVP}}$, to describe the $\pi$-electronic system coupled to nuclear degrees of freedom. This is supplemented by a Hamiltonian, $\hat{H}_{\epsilon}$, that explicitly breaks both the particle-hole and two-fold rotation symmetries of idealized carotenoid structures. The electronic degrees of freedom are treated quantum mechanically by solving the time-dependent Schr\"odinger equation using the adaptive time-dependent DMRG (tDMRG) method, while nuclear dynamics are treated via the Ehrenfest equations of motion. By defining adiabatic excited states as the eigenstates of the full Hamiltonian, $\hat{H}=\hat{H}_{\textrm{UVP}}+\hat{H}_{\epsilon}$, and diabatic excited states as eigenstates of $\hat{H}_{\textrm{UVP}}$, we present a computational framework to monitor the internal conversion process from the initial photoexcited $1^1 B_u^+$ state to the singlet triplet-pair states of carotenoids.
We further incorporate Lanczos-DMRG to the  tDMRG-Ehrenfest method to calculate transient absorption spectra from the evolving photoexcited state.  We describe in detail the accuracy and convergence criteria for DMRG, and show that this method  accurately describes the dynamical processes of  carotenoid excited states. We also discuss the effect of the symmetry breaking term, $\hat{H}_{\epsilon}$, on the internal conversion process, and show that its effect  on the extent of internal conversion can be described by a Landau-Zener-type transition. This methodological paper is a companion to our more explanatory discussion of carotenoid excited state dynamics in, \textit{Photoexcited state dynamics and singlet fission in carotenoids}, D.\ Manawadu, T.\ N.\ Georges and W.\ Barford, \emph{J. Phys. Chem. A} (2023).
\end{abstract}

\newpage

\section{Introduction}

Theoretical studies of the exotic nature of polyene excited states were pioneered by a seminal paper from Schulten and Karplus, which described the experimentally observed low-lying dark excited state of polyenes.\cite{Hudson1972,Schulten1972} They described the conjugated $\pi$-electron system using the semi-empirical Pariser-Parr-Pople (PPP) Hamiltonian\cite{Pariser1956} and the configuration interaction (CI) method description of the wavefunction with double excitations. Their work was followed by several studies based on semi-empirical Hamiltonians, which helped formulate the theoretical understanding of polyene excited states.\cite{Cizek1974,SZABO1976173,Schulten1976,Ohmine1978,Tavan1979,Lasaga1980,Tavan1987}

Early \emph{ab-initio} calculations of polyene excited states were based on self consistent field (SCF) and CI calculations.\cite{Whitten1968,hosteny1975binitio} Improvements to the ground and excited state geometries of polyenes were brought about by the use of multiconfiguration self consistent field (MCSCF) method.\cite{Aoyagi1985} Serrano-Andres and coworkers introduced the second order perturbation theory method CASPT2 with a complete active space SCF (CASSCF) wavefunction as the reference state to study electronic states of polyenes.\cite{Serrano-Andres1993a} Their calculation provided first evidence from an \emph{ab-initio} study for the existence of a dark low-lying polyene excited state. More recently, time-dependent density functional theory (TD-DFT),\cite{Hsu2001,Silva-Junior2008} extended algebraic diagrammatic construction (extended-ADC(2))\cite{Starcke2006}, CASSCF with $n$-electron valence perturbation theory (NEVPT),\cite{Angeli2011} and density functional theory with multireference configuration interaction\cite{Marian2008,Kleinschmidt2009} have been utilized to study polyene excited states. However, application of complete active space methods to model long chain polyene systems is challenging because of the exponential growth of the many-body Hilbert space with the size of the single particle basis.

In 1992, White introduced the density matrix renormalization group (DMRG) algorithm to study strongly correlated quantum lattice systems.\cite{White1992} DMRG was quickly adapted to study polyene photophysics, as applied to one-dimensional systems it yields quasi-exact results for a finite-size Hilbert space.\cite{Feiguin2013,Fano1998,PhysRevLett.82.1514,PhysRevB.63.195108,PhysRevB.65.075107,Barford2013}  DMRG was first utilized to model \emph{ab-initio} Hamiltonians in 1999.\cite{White1999} Ghosh et al. demonstrated that the CASSCF method can be incorporated to a DMRG algorithm, allowing for modelling polyene systems of natural carotenoid lengths.\cite{Ghosh2008} While the inherent multireference nature of DMRG accounts for static correlations, perturbative theory corrections are required to accurately describe the dynamic correlations present in the system.\cite{Helgaker2014,Wouters2014} More recent \emph{ab-initio} studies on polyene excited states using multireference perturbation theory (MPRT) DMRG, which accounts for the dynamic correlations have reignited the debate on carotenoid excited state energy ordering.\cite{Taffet2019e,Khokhlov2020b}


The original DMRG formulation of White has been extended to form a family of time-dependent DMRG (TD-DMRG) algorithms, designed to model time dependent phenomena of molecular systems. One of the widely used TD-DMRG algorithms is the adaptive time-dependent DMRG (tDMRG) algorithm, independently developed by Daley et al.,\cite{Daley2004a} and White and Feiguin\cite{White2004b} to study time evolution of weakly entangled systems. Examples of applications of adaptive tDMRG in molecular physics include modelling magnetization transport in spin-$\frac{1}{2}$ chains,\cite{Gobert2005a} demonstrating spin-charge separation in cold Fermi gases,\cite{Kollath2005} calculating zero temperature conductance of strongly correlated nanostructures,\cite{Al-Hassanieh2006} elucidating transport properties of quantum-dot systems connected to metallic leads\cite{DiasDaSilva2008}, evaluating spectral functions of spin-$1$ Heisenberg antiferromagnetic chain,\cite{White2008} exciton transport in one-dimensional Hubbard insulators,\cite{Al-Hassanieh2008} and non-equilibrium transport in single-impurity Anderson model.\cite{Heidrich-Meisner2009}  Techniques based on tensor network models have  recently been used to study dynamics of photophysical systems, for example, ultrafast relaxation and localization of photoexcited states in light emitting polymers\cite{Mannouch2018,Barford2018c,Perez2021}, internal conversion in pyrazine,\cite{Xie2019} and singlet fission in substituted pentacene dimers.\cite{Schroder2019} Readers are referred to a recent review by Ren and co-workers on applications of different TD-DMRG algorithms to model dynamics of quantum systems.\cite{Ren2022}

While state-of-the-art \emph{ab-initio} methods have had great success in calculating static properties of polyene excited states, due to computational expediencies the use of semi-empirical Hamiltonians with a single electron basis is more attractive to model dynamical processes of photoexcited polyenes. With the correct parametrization DMRG has been shown to work very well for semi-empirical Hamiltonians with a reduced single-particle basis.\cite{Chandross1997,Castleton2002} The DMRG algorithm with the PPP Hamiltonian has been widely used to study the electronic properties of conjugated polyenes.\cite{Yaron1998,Fano1998,PhysRevLett.82.1514,PhysRevB.63.195108,PhysRevB.65.075107,PhysRevB.66.205112,doi:10.1063/1.3149536}


Recently, the present authors developed DMRG methods to simulate internal conversion of photoexcited states and singlet triplet-pair production in carotenoid systems of up to $22$ conjugated carbon atoms. Carotenoids are particularly challenging, as they exhibit strong electronic correlation and strong electron-nuclear coupling. In dimers, they also exhibit singlet fission after photoexcitation.
In ref \citen{Manawadu2022} we implemented mixed quantum-classical dynamics, treating the electrons via adaptive tDMRG and the nuclei via Ehrenfest dynamics, to simulate internal conversion in zeaxanthin. In a companion paper\cite{Manawadu2022b} we extend those simulations to neurosporene and we also describe our calculations of transient absorption.

The primary purpose of this paper is to describe in more detail the methodology of the tDMRG-Ehrenfest simulation for a wider theoretical chemistry community. In particular, we emphasize that tDMRG is a rather natural generalization of the finite-lattice algorithm of static DMRG.
We also explain the Lanczos-DMRG method for computing the transient absorption spectrum of the time-evolving photoexcited state. We show that DMRG methods can accurately and reliably describe the complex photoexcited state dynamics of large linear conjugated systems.

In this paper we also explore in more detail the excited state dynamics as a function the broken-symmetry perturbation that connects the diabatic states of opposite particle-hole symmetry. We show that for a small perturbation, the system undergoes a  transition from the initial adiabatic state, $S_1$, to $S_2$, while remaining in the same diabatic state, $1^1 B_u^+$. In contrast, for larger perturbations, the system evolves adiabatically  on the $S_1$ surface, changing character from the excitonic $1^1 B_u^+$ state to the singlet triplet-pair $2^1 A_g^-$ state. In all cases the $S_1$ and $S_2$ energies exhibit an avoided crossing, and the dynamics can be approximately modeled as a two-level system.


\section{Model Hamiltonian and eigenstates}

\subsection{UV-Peierls Hamiltonian}

The PPP Hamiltonian, routinely utilized with static DMRG for modelling electronic properties of conjugated polymers, contains long range Coulomb interactions and therefore is not readily suitable for dynamical simulations using the adaptive time-dependent DMRG (tDMRG) algorithm. In a tDMRG setting, the electronic degrees of freedom are conveniently described by the extended Hubbard (UV) Hamiltonian, defined by
\begin{equation}
	\hat{H}_{\textrm{UV}} = -2\beta \sum_{n=1}^{N-1} \hat{T}_n + U \sum_{n=1}^N  \big( \hat{N}_{n \uparrow} - \frac{1}{2} \big) \big( \hat{N}_{n \downarrow} - \frac{1}{2} \big)
+ \frac{1}{2} \sum_{n=1}^{N-1} V \big( \hat{N}_n -1 \big) \big( \hat{N}_{n+1} -1 \big), \label{eqn:UV}
\end{equation}
where  $n$ labels  the  $n^{\textrm{th}}$ C-atom, $N$ is the number of conjugated C-atoms and $N/2$ is the number of double-bonds.
$\hat{T}_ n = \frac{1}{2}\sum_{\sigma} \left( \hat{c}^{\dag}_{n , \sigma} \hat{c}_{n + 1 , \sigma} + \hat{c}^{\dag}_{n+ 1 , \sigma} \hat{c}_{n  , \sigma} \right)$  is the bond order operator, $\hat{c}^{\dag}_{n , \sigma} (\hat{c}^{}_{n , \sigma})$ creates (destroys) an electron with spin $\sigma$ in the $p_z$ orbital of the $n^{\textrm{th}}$ C-atom, and $\hat{N}_n$ is the number operator. $U$ and $V$ correspond to Coulomb parameters which describe interactions of two electrons in the same orbital and nearest neighbors respectively, and $\beta$ represents the electron hopping integral between neighboring C-atoms. The UV-Peierls Hamiltonian is invariant to a particle-hole transformation. For idealized carotenoid structures
with $C_{2}$ symmetry, its eigenstates will have definite $C_{2}$ and particle-hole symmetries.\cite{Barford2013}

With the inclusion of nuclear degrees of freedom, the UV-Peierls Hamiltonian is defined by
\begin{equation}
	\hat{H}_{\textrm{UVP}} = \hat{H}_{\textrm{UV}} + 2 \alpha \sum_{n=1}^{N-1}  \left( u_{n+1} - u_n \right)  \hat{T}_n + \frac{K}{2} \sum_{n=1}^{N-1} \left( u_{n+1} - u_n \right)^2 , \label{eqn:UVP}
\end{equation}
where $\alpha$ is the electron-nuclear  parameter, $K$ is the nuclear spring constant, and $u_n$ is the displacement of the $n^{\textrm{th}}$ carbon atom from its undistorted geometry. Electron hopping integrals relate to nuclear geometries via $\beta_n = \beta - \alpha ( u_{n+1} - u_n ).$

\subsection{Broken symmetry}

\begin{figure}
		\includegraphics[width=0.9\textwidth,height=\textheight,keepaspectratio]{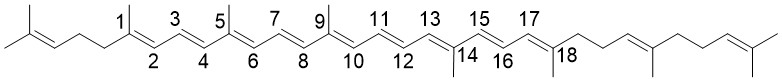}
		\caption{Neurosporene, a naturally occurring carotenoid with $18$ conjugated carbon atoms.}
		\label{Figure1}
	\end{figure}

In order to facilitate internal conversion from the photoexcited $1^1 B_u^+$ state to the triplet-pair states (which are of negative particle-hole character), an interaction term which breaks the particle-hole symmetry is introduced.\footnote{We follow the particle-hole sign convention of ref \citen{Valentine20}, which is commonly used by the experimental community but is the opposite definition to refs \citen{PhysRevB.63.195108} and \citen{Barford2013}.} The symmetry breaking term is defined by
\begin{equation}
	\hat{H}_\epsilon
		= \sum_{n=1}^{N} \epsilon_n (\hat{N}_n -1), \label{eqn:HSB}
\end{equation}
where $\epsilon_n$ is the on-site potential energy parameter on the $n^{\textrm{th}}$ C-atom. The inclusion of $H_{\epsilon}$ is justified as carotenoids with both electron donating and withdrawing substituent groups (e.g., neurosporene shown  in Figure \ref{Figure1}), which act as electron donors and acceptors to the $\pi$-system, are known to undergo singlet fission.\cite{Hashimoto2018}

Since we are only interested in the singlet excitations, we project out the high spin contributions to the Hilbert space by supplementing the Hamiltonian with
\begin{equation}
	\hat{H}_\lambda = \lambda \hat{S}^2,
\end{equation}
where $\hat{S}$ is the total spin operator, and $\lambda > 0$.

We now define the full Born-Oppenheimer Hamiltonian as
\begin{equation}
	\hat{H} = \hat{H}_{\textrm{UVP}} + \hat{H}_\lambda + \hat{H}_\epsilon.
\end{equation}
Eigenstates of $\hat{H}$ are labelled adiabatic (singlet) eigenstates, and are defined by
\begin{equation}
	\hat{H} \ket{S_i} = E_i \ket{S_i}.
\end{equation}
We define a diabatic (singlet) basis spanned by the eigenstates of ($\hat{H}_{\textrm{UVP}}+\hat{H}_\lambda$) as
\begin{equation}
	\left (\hat{H}_{\textrm{UVP}} + \hat{H}_\lambda \right ) \ket{\phi_j} = E_{\phi_j} \ket{\phi_j}.
\end{equation}
The diabatic eigenstates  have definite $C_{2}$ and particle-hole symmetries.

We  calculate the probability that the system described by $\Psi(t)$ occupies the adiabatic state $S_i$ by
\begin{equation}
	P(\Psi(t); S_i) = |\braket{S_i}{\Psi(t)}|^2 \label{eq:popad}
\end{equation}
and the probability that it occupies the diabatic state $\phi_j$ by
\begin{equation}
	P(\Psi(t); \phi_j) = |\braket{\phi_j}{\Psi(t)}|^2. \label{eq:popdia}
\end{equation}
Finally,  the probabililty that the adiabat $S_i$ occupies the diabat $\phi_j$ is
\begin{equation}
	P(S_i; \phi_j) = |\braket{\phi_j}{S_i}|^2. \label{eq:projection}
\end{equation}

For our simulations, we first determine the ground state of the system using the static DMRG method by solving eq \eqref{eqn:UVP} for fixed nuclear displacements, $\{u_n\}$. Starting from the undimerised geometry ($u_n=0, \forall n$), ground and excited state energies and geometries are found by iterative application of eq \eqref{eqn:fn} with the force per atom $n$, $f_\mathrm{n}=0$.\cite{PhysRevB.63.195108} The initial system, $\Psi(t=0)$, is taken to be the adiabatic singlet state $S_i$ with the largest projection onto the $1^1 B_u^+$ diabat, determined via eq \eqref{eq:projection}. This corresponds to a dipole-allowed vertical excitation from the ground state, $S_0$. The choice to define the initial system this way, instead of as $\hat{\mu} S_0$, is based on the observation that the Ehrenfest approximation (discussed in section \ref{Ehr}) is most accurate for systems evolving on a single adiabatic potential energy surface.

Diabatic eigenstates of $\hat{H} = (\hat{H}_{\textrm{UVP}} + \hat{H}_\lambda)$ with a positive particle-hole symmetry are termed `ionic', as the expectation value of the ionicity operator, $\hat{I} = \sum_n \left(\hat{N}_n - 1 \right)^2$, is larger for these states than for eigenstates  with a negative particle-hole symmetry, termed `covalent'. We use this property of a larger ionicity for ionic states to distinguish them from covalent states during the dynamical simulation.

\subsection{Parametrizations}

\subsubsection{UVP Model parametrization}

The UV Hamiltonian does not contain the long-range Coulomb interactions of the PPP Hamiltonian, and therefore requires a parametrization of the $U$ and $V$ Coulomb parameters to replicate the PPP model predictions. While retaining the Chandross-Mazumdar parametrization of $\beta = 2.4$ eV,\cite{Chandross1997} $K = 46$ eV {\AA}$^{-2}$ and $\alpha = 4.593$ eV {\AA}$^{-1}$ from Barford and co-workers,\cite{PhysRevB.63.195108} in our earlier work\cite{Manawadu2022} we parametrized the UV model for internal conversion from $1^1 B_u^+$ to $1^1 B_u^-$ to reproduce the predictions of ref \citen{Valentine20}.

\begin{figure}[h!]
 \includegraphics[height=0.8\textheight]{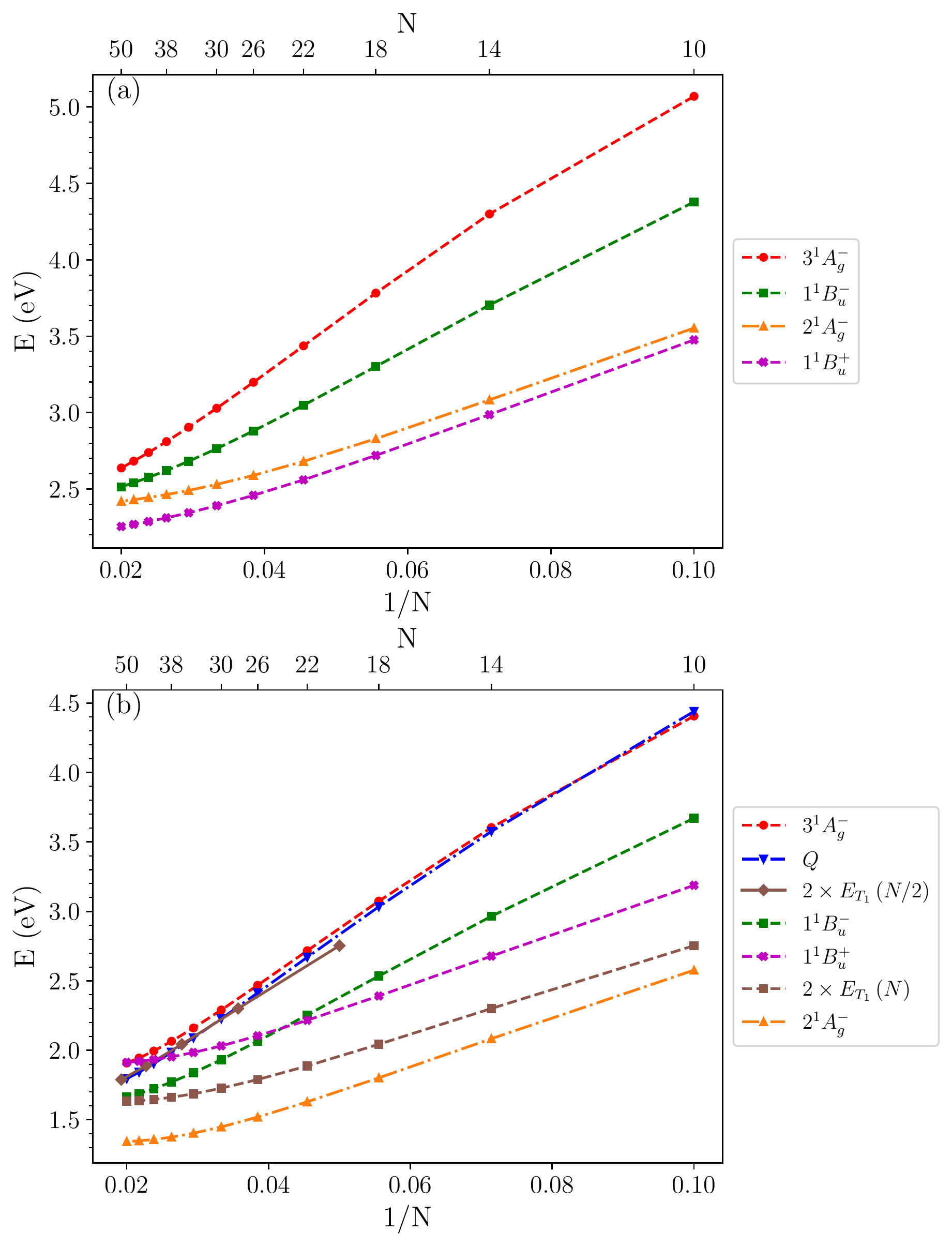}
  \caption{Vertical (a) and relaxed (b) singlet excitation energies of the UV-Peierls model with $U = 7.25$ eV and $V = 3.25$ eV, found by solving eq \eqref{eqn:UVP}. $N$ is the number of conjugated carbon atoms of the system. The vertical energy gaps of $\sim 0.1$ eV between $1^1 B_u^+$(magenta) and $2^1 A_g^-$(orange) for $18 \le N \le 26$ agree with the corresponding excitation energies reported in ref \cite{Taffet2019e}.}
  \label{Figure2}
\end{figure}

In our companion paper\cite{Manawadu2022b}, we model internal conversion from the $1^1 B_u^+$ state to both the $1^1 B_u^-$ and $2^1 A_g^-$ states. For the latter, we require a parametrization where $E_{1^1 B_u^+}$(vertical)  $< E_{2^1 A_g^-}$(vertical). For a given $U$, increasing $V$ decreases ($E_{1^1 B_u^+} - E_{2^1 A_g^-}$). Keeping all other parameters the same (i.e., $U=7.25$ eV and $\beta=2.4$ eV), we find $V=3.25$ eV such that $E_{1^1 B_u^+}$(vertical)  $- E_{2^1 A_g^-}$(vertical) replicates the lowest-lying carotenoid dark and bright state vertical excitation energies reported in Table 2 of ref \citen{Taffet2019e}. The diabatic vertical and relaxed energies for the UV-Peierls model with these parameters are illustrated in Figure \ref{Figure2}. (The corresponding figures where $E_{1^1 B_u^+}$(vertical)  $> E_{2^1 A_g^-}$(vertical) for $U=7.25$ eV, $\beta=2.4$ eV, and $V=2.75$ eV are shown in ref \citen{Manawadu2022}). For all carotenoid chain lengths under consideration, $1^1 B_u^+$ vertical energies lie below $2^1 A_g^-$ vertical energies, while $1^1 B_u^+$ relaxed energies are above $2^1 A_g^-$ relaxed energies, indicating the possibility of internal conversion from  the $1^1 B_u^+$ to the $2^1 A_g^-$ states via a diabatic energy level crossing. Internal conversion to the $2^1 A_g^-$ state could potentially lead to endothermic intramolecular singlet fission, as the relaxed energy of the $2^1 A_g^-$ state is lower than the energy of two triplets on the same chain. However, as shown in ref \citenum{Manawadu2022b}, exothermic intermolecular singlet fission is possible provided that the carotenoids are twisted in their ground state.

\subsubsection{Parametrizing $\hat{H}_{\epsilon}$}

%
%

The symmetry breaking term (eq \eqref{eqn:HSB}) alters the on-site potential energies, and therefore changes the Mulliken charge densities of the $\pi$-system from unity. As outlined below, we use the ground state Mulliken charge densities of the $\pi$-system to parametrize $H_{\epsilon}$ for neurosporene, whose structural formula is illustrated in Figure \ref{Figure1}. The optimum Mulliken charge densities are calculated using the ORCA program package.\cite{Neese12,Neese17} Geometry optimizations are performed using density functional theory (DFT) with a B3LYP functional\cite{Stephens94} and a def2-TZVP basis set,\cite{Weigend05,Weigend06} followed by calculations of electron densities. To enforce charge neutrality in our model, the mean shifted Mulliken charge densities are used as target densities in an optimization algorithm to determine $\hat{H}_{\epsilon}$.

Allowing unconstrained optimization of $\hat{H}_{\epsilon}$ leads to unphysical on-site potential energies and significant changes in the character of excited states. In order to avoid large perturbations, we use projected gradient descent algorithm to search for $\boldsymbol{\epsilon}=\left ( \epsilon_1 , \epsilon_2 \dots \epsilon_N \right)$ such that $|\epsilon_i| < \epsilon_{\textrm{max}}$, $\forall \ i \in \{1,2, \dots , N\}$.\cite{LEVITIN19661} For a given $\boldsymbol{\epsilon}$, $\boldsymbol{d} = ( d_1, d_2, \dots , d_N)$ where $d_i = \expval{\hat{N}_i - 1}{\Psi}$ can be found via the static DMRG algorithm. We define the minimization function as $E(\boldsymbol{\epsilon}) = \lVert \boldsymbol{d_\textrm{opt}} - \boldsymbol{d} (\boldsymbol{\epsilon}) \rVert$ where $\boldsymbol{d_\textrm{opt}}$ is the target density vector found via DFT. The algorithm is as follows:

\begin{enumerate}
\item Choose initial $\boldsymbol{\epsilon}_0$ within the constraints
\item Loop until the convergence condition is met:
\begin{enumerate}
\item Find the descent direction $-\nabla  E(\boldsymbol{\epsilon}_k)$
\item Find $\bar{\boldsymbol{\epsilon}}_{k+1} = \boldsymbol{\epsilon}_k -\nabla  E(\boldsymbol{\epsilon}_k)$
\item Projection: Find $\boldsymbol{\epsilon}_{k+1} = \left ( \epsilon^{k+1}_1 , \epsilon^{k+1}_2 \dots \epsilon^{k+1}_N \right)$ such that $\forall \ i \in \{1,2, \dots , N\}$,
\begin{equation}
 \epsilon^{k+1}_i =
\begin{cases}
      \bar{\epsilon}^{k+1}_i &,  |\bar{\epsilon}^{k+1}_i| \leq \epsilon_{\textrm{max}} \\
      \epsilon_{\textrm{max}} &,  \bar{\epsilon}^{k+1}_i > \epsilon_{\textrm{max}} \\
      -\epsilon_{\textrm{max}} &,  \bar{\epsilon}^{k+1}_i < -\epsilon_{\textrm{max}}
   \end{cases}
\label{eqn:CGD}
\end{equation}
\end{enumerate}
\item Convergence is evaluated via the coefficient of variation $r^2 (\boldsymbol{\epsilon}) $ defined as:
\begin{equation}
r^2  (\boldsymbol{\epsilon})= 1 - \frac{ \left[E(\boldsymbol{\epsilon}) \right ]^2 }
{\lVert \boldsymbol{d_\textrm{opt}} \rVert^2}
\label{eqn:similarity}
\end{equation}

\end{enumerate}


\begin{table}[h]
    \centering
    \begin{tabular}{|r|r|r|r|}
    \hline
        Carbon site, $n$ & Mulliken charges (q) & $\epsilon_n$ (eV) & $\expval{\hat{N}_n -1}$ \\ \hline
        1 & 0.14 & -1.00 & 0.17 \\ \hline
        2 & -0.18 & 0.56 & -0.14 \\ \hline
        3 & -0.05 & 1.00 & 0.06 \\ \hline
        4 & -0.18 & 0.82 & -0.10 \\ \hline
        5 & 0.15 & -1.00 & 0.15 \\ \hline
        6 & -0.14 & 0.02 & -0.10 \\ \hline
        7 & -0.07 & 1.00 & 0.03 \\ \hline
        8 & -0.16 & 0.84 & -0.09 \\ \hline
        9 & 0.13 & -1.00 & 0.12 \\ \hline
        10 & -0.09 & -0.01 & -0.07 \\ \hline
        11 & -0.11 & 1.00 & -0.04 \\ \hline
        12 & -0.10 & 1.00 & -0.03 \\ \hline
        13 & -0.11 & 0.09 & -0.09 \\ \hline
        14 & 0.14 & -1.00 & 0.13 \\ \hline
        15 & -0.18 & 0.92 & -0.10 \\ \hline
        16 & -0.05 & 1.00 & 0.06 \\ \hline
        17 & -0.19 & 0.61 & -0.14 \\ \hline
        18 & 0.14 & -1.00 & 0.17 \\ \hline
    \end{tabular}
    \caption{The $\pi$-electron Mulliken charges from the \emph{ab-initio} DFT calculation, parameters for the symmetry-breaking term, $\hat{H}_{\epsilon}$, and the expectation values of number densities calculated from the parametrized $\hat{H}_{\epsilon}$. In order to maintain $\pi$-electron charge neutrality, each \emph{ab-initio} charge was increased by $0.05q$. The chemical formula of neurosporene is shown in figure \ref{Figure1}.}
    \label{Heps}
\end{table}

We perform the optimization for $\epsilon_{\textrm{max}}=1.0$ eV. For the simulations described in ref \citen{Manawadu2022b}, $\epsilon_{\textrm{max}}$ is constrained to an upper bound of $1.0$ eV to prevent the formation of an unphysical potential energy gradients across the conjugated carbon atoms which causes an unphysical mixing of the ionic and covalent states. (However, the effect of an arbitrary symmetry-breaking potential is described in section \ref{Se:7.2} of this paper.) The optimized $\hat{H}_{\epsilon}$ found for neurosporene with $V=3.25$ eV and $r^2 (\boldsymbol{\epsilon}) = 0.92$ is shown in Table \ref{Heps}.



%
%
%
%
%
%
%


\section{Density Matrix Renormalization Group (DMRG)}

From now on in this paper we define a `site' as  a $p_z$ orbital of a C-atom. The single-site  basis for the UVP model defined in eq (\ref{eqn:UVP}), i.e.,  \{
\begin{tikzpicture}[line width=0.5pt,baseline={([yshift=-9pt]current bounding box.north)}]
    \draw (0,0) -- (0.4cm,0);
    \node[] at (0.5cm,0) {,};
    \begin{scope}[xshift=0.6cm]
    \draw (0,0) -- (0.4cm,0);
    \node[] at (0.5cm,0) {,};
    \draw [arrows = {-Straight Barb[left,scale=0.5]}] (1ex,-1ex) -- (1ex,1ex);
    \end{scope}
    \begin{scope}[xshift=1.2cm]
    \draw (0,0) -- (0.4cm,0);
        \node[] at (0.5cm,0) {,};
     \draw [arrows = {-Straight Barb[left,scale=0.5]}] (1ex,1ex) -- (1ex,-1ex);
    \end{scope}
    \begin{scope}[xshift=1.8cm]
   \draw (0,0) -- (0.4cm,0);
     \draw [arrows = {-Straight Barb[left,scale=0.5]}] (1ex,-1ex) -- (1ex,1ex);
    \draw [arrows = {-Straight Barb[left,scale=0.5]}] (1.4ex,1ex) -- (1.4ex,-1ex);
    \end{scope}
  \end{tikzpicture}
      \}, has a dimensionality of $4$. Therefore, exactly solving the time-dependent Schr\"odinger equation for $N=18$, the relevant carotenoid chain length, would require solving a $S_z=0$ Hilbert space of size $\approx 2.4 \times 10^{9}$. This is not feasible in realistic time scales. DMRG methods are based on the premise that by an efficient truncation of the exact Hilbert space to retain only the important many-particle states, the most important features of the system can be preserved at a significantly lower computational cost. We begin this section by describing the static DMRG algorithm and then show how the method can easily be extended to the adaptive time-dependent DMRG (tDMRG) algorithm.

\subsection{Static DMRG algorithm}\label{SDMRG}

\begin{figure}
{
\begin{tikzpicture}[
    font=\sffamily,
    level/.style={black,thick},
    sublevel/.style={black,densely dashed},
    ionization/.style={black,dashed},
    transition/.style={black,->,>=stealth',shorten >=1pt},
    radiative/.style={transition,decorate,decoration={snake,amplitude=1.5}},
    indirectradiative/.style={radiative,densely dashed},
    nonradiative/.style={transition,dashed},
    thick,scale=0.6, every node/.style={scale=0.6},
  ]

%
%
%

  \tikzstyle{every node}=[font=\small]
\coordinate (A1) at (-5,10);
\coordinate (A2) at (-1,9);
\coordinate (D1) at (0,10);
\coordinate (D2) at (4,9);

  \draw[fill=red] (A1) rectangle (A2);
  \draw[fill=green] (D1) rectangle (D2);
  \draw[->] (-3.5,8.9) -- (-3.5,7.1);
    \draw[->] (2.5,8.9) -- (2.5,7.1);

  \node[align=center,font=\small] at (-3,10.5) {block S};
    \node[align=center,font=\small] at (2,10.5) {block E};

    \node[align=center,font=\small] at (-0.5,7.5) {two sites};
    \node[align=center,font=\small] at (-3,9.5) {$k-1$ sites};
        \node[align=center,font=\small] at (2,9.5) {$k-1$ sites};

     \draw[->] (-4.5,5.9) -- (-4.5,4.1);
    \draw[->] (3.5,5.9) -- (3.5,4.1);
        \draw[->] (-1.5,5.9) -- (-1.5,5.1);
                \draw[->] (0.5,5.9) -- (0.5,5.1);

        \node[align=center,font=\small] at (-0.5,4.5) {superblock};

	 \node[align=center,font=\small] at (-4,1.5) {new block S};
    \node[align=center,font=\small] at (3,1.5) {new block E};

     \draw[->] (-6.5,2.9) -- (-6.5,1.1);
    \draw[->] (5.5,2.9) -- (5.5,1.1);





   \coordinate (A1) at (-7,7);
\coordinate (A2) at (-3,6);
\coordinate (B) at (-1.5,6.5);
\coordinate (C) at (0.5,6.5);
\coordinate (D1) at (2,7);
\coordinate (D2) at (6,6);

  \draw[fill=red] (A1) rectangle (A2);
  \draw (B) circle (0.5cm);
  \draw (C) circle (0.5cm);
  \draw[fill=green](D1) rectangle (D2);

    \coordinate (P) at (-1,6);
	\node [fit=(A1) (P),draw,dotted,blue] {};

\coordinate (Q) at (0,7);
	\node [fit=(Q) (D2),draw,dotted,blue] {};


%


  \coordinate (A1) at (-7,4);
\coordinate (A2) at (-0.5,3);
\coordinate (D1) at (-0.5,4);
\coordinate (D2) at (6,3);

  \draw[fill=red] (A1) rectangle (A2);
  \draw[fill=green] (D1) rectangle (D2);

%
%
%


  \coordinate (A1) at (-7,1);
\coordinate (A2) at (-1,0);
\coordinate (B) at (3.5,0.5);
\coordinate (C) at (5.5,0.5);
\coordinate (D1) at (0,1);
\coordinate (D2) at (6,0);

  \draw[fill=red] (A1) rectangle (A2);
  \draw[fill=green] (D1) rectangle (D2);
      \node[align=center,font=\small] at (-4,0.5) {$k$ sites};
            \node[align=center,font=\small] at (3,0.5) {$k$ sites};

%
%
%

\end{tikzpicture}
	\caption{Schematic diagram illustrating one iteration of the infinite DMRG algorithm.\cite{Schollwock2005}}
	\label{Figure3}
}
\end{figure}
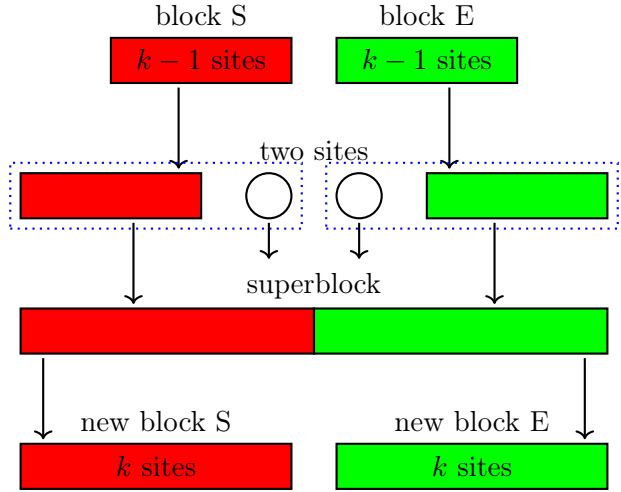

The infinite DMRG algorithm was introduced in 1992 to accurately calculate ground states of one-dimensional quantum systems.\cite{White1992} Suppose that a system of length ($k-1$) described by a Hilbert space of size $M_l$ is spanned by the basis states $\{ \ket{l_{k-1}} \}$. Consider the process of the linear growth of this system block by adding a single site at index $k$ (see Figure \ref{Figure3}). The single site is fully described by the $d$-dimensional basis $\{ \ket{\alpha_{k} } \}$. An augmented system block of length $k$ is constructed in the product Hilbert space spanned by $\{ \ket{l_{k-1}} \otimes \ket{\alpha_{k} } \}$, with dimensions $N_S =M_l \times d$. An analogous augmented environment is constructed in the product Hilbert space spanned by $\{ \ket{\alpha_{k+1}} \otimes \ket{r_{k-1}}    \}$, with dimensions $N_E =M_r \times d$. A superblock of length $2k$ is now formed in the product Hilbert space spanned by $\{ \ket{l_{k-1}} \otimes \ket{\alpha_{k} } \otimes \ket{\alpha_{k+1}} \otimes \ket{r_{k-1}} \}$. The ground state,
 \begin{equation}
	\ket{\Psi} =
	\sum_{l_{k-1}=1}^{M_l}
	\sum_{\alpha_{k}=1}^{d}
	\sum_{\alpha_{k+1}=1}^{d}
	\sum_{r_{k-1}=1}^{M_r}
	\Psi_{l_{k-1},\alpha_{k},\alpha_{k+1},r_{k-1}}
	\ket{l_{k-1}} \ket{\alpha_{k}} \ket{\alpha_{k+1}} \ket{r_{k-1}},
\end{equation}
is obtained by a sparse-matrix diagonalization (e.g., conjugate gradient or Davidson) of the Hamiltonian in the superblock basis.

Defining the augmented system block state $\ket{i} \in \{ \ket{l_{k-1}} \otimes \ket{\alpha_{k} } \}$, and the augmented environment block state $\ket{j} \in \{ \ket{\alpha_{k+1}} \otimes \ket{r_{k-1}}    \}$, the ground state may also be expressed as,
 \begin{equation}
	\ket{\Psi} =
	\sum_{i=1}^{N_S}
	\sum_{j=1}^{N_E}
	\Psi_{i,j}
	\ket{i} \ket{j}.
	\label{eqn:augblock}
\end{equation}
Now a truncation procedure must be introduced to describe the system block of size $k$ using a basis of dimension $M_S < N_S$. Suppose that the ground state of the system can be expressed by the approximate state $\tilde{\ket{\Psi}}$ in this truncated Hilbert space,
 \begin{equation}
	\ket{\tilde{\Psi}} =
	\sum_{p=1}^{M_S}
	\sum_{j=1}^{N_E}
	\tilde{\Psi}_{p,j}
	\ket{p} \ket{j}.
\end{equation}
Finding the optimum $\ket{\tilde{\Psi}}$ is achieved by the minimization of the quadratic norm $S_2$,
\begin{equation}
S_2= \lVert \ket{\Psi} - \ket{\tilde{\Psi}} \rVert^2. \label{eqn:S}
\end{equation}

From eq \eqref{eqn:augblock}, we see that $\Psi$ can be recast into a rectangular matrix  of dimension $N_S \times N_E$, which can then be decomposed  using singular value decomposition as
\begin{equation}
	\Psi = U D V^\dagger,
\end{equation}
where $U$ is a unitary matrix of dimension $N_S \times N_E$, $V$ is a unitary matrix of dimension $N_E \times N_E$, and $D$ is a diagonal matrix of dimension $N_E \times N_E$ with elements $\{\lambda_\beta\}$.
This transformation implies that $\ket{\Psi}$ can be expressed as a Schmidt decomposition,
\begin{equation}
	\ket{\Psi}
		=
			 \sum_{\beta=1}^r \lambda_\beta \ket{\beta_S} \ket{\beta_E}, \label{eqn:Schm}
\end{equation}
where $\ket{\beta_S}=\sum_i U_{i,\beta} \ket{i}$, $\ket{\beta_E}=\sum_j V_{\beta,j}^* \ket{j}$, and $r=\min(N_S,N_E)$.\cite{Feiguin2013} It follows that in the Schmidt basis, the reduced density operator $\hat{\rho}_S = \trace_E \ket{\Psi}\bra{\Psi}$ can be written as
\begin{equation}
	\hat{\rho}_S = \sum_{\beta=1}^{N_S} \lambda^2_\beta \ket{\beta_S} \bra{\beta_S}, \label{eqn:Schmidt}
\end{equation}
where $\ket{\beta_S}$ and $\lambda^2_\beta = \omega_\beta$ are  the eigenstates and eigenvalues, respectively, of the reduced density operator.

The quadratic norm $S_2$ is given by
\begin{equation}
	S_2
		=
			\sum_{\beta=1}^{r} \omega_\beta - \sum_{\beta=1}^{M_S} \omega_\beta
		=
			\sum_{\beta=M_S+1}^{r} \omega_\beta . \label{eqn:S2}
\end{equation}
$S_2$ is therefore minimized by retaining the $M_S$ eigenstates of $\hat{\rho}_S$ with the largest eigenvalues.

Once the truncated basis for the new system block of length $k$, $\{\ket{l_{k}} \}$ is known, all operators, including the Hamiltonian, are rotated to the new Hilbert space via a similarity transformation. Suppose $\hat{O}=\sum_{n=1}^k \hat{O}_n$ is a generic operator. Then $\hat{O}_k$ is given by
\begin{equation}
	\mel{l_{k}}{\hat{O}_k}{\tilde{l}_{k}}
		=
			\sum_{l_{k-1}=1}^{M_l}
			\sum_{\alpha_{k}=1}^{d}
			\sum_{\tilde{\alpha}_{k}=1}^{d}
				\bra{l_{k}} \ket{l_{k-1}}\otimes \ket{\alpha_{k}}
				\mel{\alpha_{k}}{\hat{O}_k}{\tilde{\alpha}_{k}}
				\bra{l_{k-1}}\otimes \bra{\tilde{\alpha}_{k}} \ket{\tilde{l}_{k}}.
\end{equation}
$\hat{O}_n$ where $n<k$ is transformed to the truncated Hilbert space by
\begin{equation}
	\mel{l_{k}}{\hat{O}_n}{\tilde{l}_{k}}
		=
			\sum_{l_{k-1}=1}^{M_l}
			\sum_{\tilde{l}_{k-1}=1}^{M_l}
			\sum_{\alpha_{k}=1}^{d}
				\bra{l_{k}} \ket{l_{k-1}}\otimes \ket{\alpha_{k}}
				\mel{l_{k-1}}{\hat{O}_n}{\tilde{l}_{k-1}}
				\bra{\tilde{l}_{k-1}}\otimes \bra{\alpha_{k}} \ket{\tilde{l}_{k}}.
\end{equation}

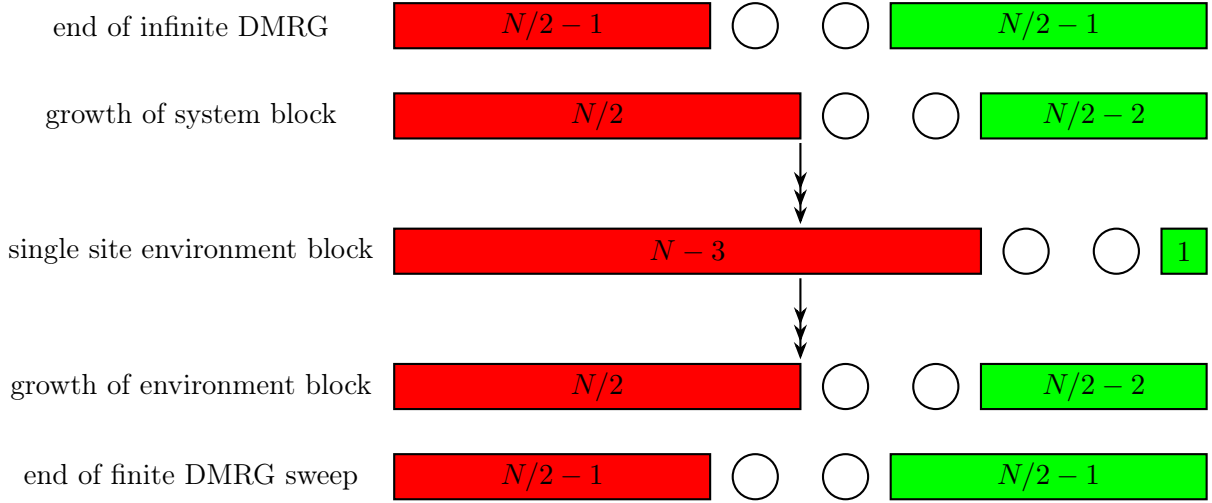
\begin{figure}[h!]
{
\begin{tikzpicture}[
    font=\sffamily,
    level/.style={black,thick},
    sublevel/.style={black,densely dashed},
    ionization/.style={black,dashed},
    transition/.style={black,->,>=stealth',shorten >=1pt},
    radiative/.style={transition,decorate,decoration={snake,amplitude=1.5}},
    indirectradiative/.style={radiative,densely dashed},
    nonradiative/.style={transition,dashed},
    thick,scale=0.6, every node/.style={scale=0.6},
  ]

%
%
%

  \tikzstyle{every node}=[font=\small]

  \node[align=center,font=\small] at (-14.5,8.5) {end of infinite DMRG};
   \coordinate (A1) at (-10,9);
\coordinate (A2) at (-3,8);
\coordinate (B) at (-2,8.5);
\coordinate (C) at (0,8.5);
\coordinate (D1) at (1,9);
\coordinate (D2) at (8,8);

  \draw[fill=red] (A1) rectangle (A2);
  \draw (B) circle (0.5cm);
  \draw (C) circle (0.5cm);
  \draw[fill=green](D1) rectangle (D2);


     \node[align=center,font=\small] at (-14.5,6.5) {growth of system block};
   \coordinate (A1) at (-10,7);
\coordinate (A2) at (-1,6);
\coordinate (B) at (-0,6.5);
\coordinate (C) at (2,6.5);
\coordinate (D1) at (3,7);
\coordinate (D2) at (8,6);

  \draw[fill=red] (A1) rectangle (A2);
  \draw (B) circle (0.5cm);
  \draw (C) circle (0.5cm);
  \draw[fill=green](D1) rectangle (D2);

\node[align=center,font=\small] at (4.5,8.5) {$N/2-1$};
\node[align=center,font=\small] at (-6.5,8.5) {$N/2-1$};


 \node[align=center,font=\small] at (5.5,6.5) {$N/2-2$};
\node[align=center,font=\small] at (-5.5,6.5) {$N/2$};

   \draw [-{Stealth}{Stealth}{Stealth}] (-1,5.9)--(-1,4.1);

  \coordinate (X1) at (-4,6.5);
    \coordinate (Y1) at (-1.5,6.5);




       \node[align=center,font=\small] at (-14.5,3.5) {single site environment block};

  \coordinate (A1) at (-10,4);
\coordinate (A2) at (3,3);
\coordinate (B) at (4,3.5);
\coordinate (C) at (6,3.5);
\coordinate (D1) at (7,4);
\coordinate (D2) at (8,3);

  \draw[fill=red] (A1) rectangle (A2);
  \draw (B) circle (0.5cm);
  \draw (C) circle (0.5cm);
  \draw[fill=green] (D1) rectangle (D2);

    \node[align=center,font=\small] at (7.5,3.5) {$1$};
\node[align=center,font=\small] at (-3.5,3.5) {$N-3$};


   \draw [-{Stealth}{Stealth}{Stealth}] (-1,2.9)--(-1,1.1);

    \coordinate (P1) at (3.5,1.2);
   \coordinate (P2) at  (5.5,1.2);


         \node[align=center,font=\small] at (-14.5,0.5) {growth of environment block};

   \coordinate (A1) at (-10,1);
\coordinate (A2) at (-1,0);
\coordinate (B) at (-0,0.5);
\coordinate (C) at (2,0.5);
\coordinate (D1) at (3,1);
\coordinate (D2) at (8,0);

  \draw[fill=red] (A1) rectangle (A2);
  \draw (B) circle (0.5cm);
  \draw (C) circle (0.5cm);
  \draw[fill=green](D1) rectangle (D2);

   \node[align=center,font=\small] at (5.5,0.5) {$N/2-2$};
\node[align=center,font=\small] at (-5.5,0.5) {$N/2$};



   \node[align=center,font=\small] at (-14.5,-1.5) {end of finite DMRG sweep};

   \coordinate (A1) at (-10,-1);
\coordinate (A2) at (-3,-2);
\coordinate (B) at (-2,-1.5);
\coordinate (C) at (0,-1.5);
\coordinate (D1) at (1,-1);
\coordinate (D2) at (8,-2);

  \draw[fill=red] (A1) rectangle (A2);
  \draw (B) circle (0.5cm);
  \draw (C) circle (0.5cm);
  \draw[fill=green](D1) rectangle (D2);

  \node[align=center,font=\small] at (4.5,-1.5) {$N/2-1$};
\node[align=center,font=\small] at (-6.5,-1.5) {$N/2-1$};

\end{tikzpicture}

	\caption{Schematic diagram for the finite DMRG algorithm illustrating a single finite lattice sweep.\cite{Schollwock2005} The number of sites in each block is shown as text. An open circle represents a single site block.}
	\label{Figure4}
}
\end{figure}

The total Hamiltonian is not known during the intermediate steps of the infinite DMRG algorithm, and this leads to errors, especially in systems with strong physical effects from impurities or randomness in the Hamiltonian.\cite{Schollwock2005} These finite size effects can be resolved by performing finite `sweeps' after the infinite DMRG. Once the desired system size $N$ is reached, the steps of infinite DMRG is continued, but with one block (system) growing at the expense of the other (environment). The superblock size remains fixed at $N$, and truncation of the basis is only performed for the growing block. Determination of the superblock ground state is efficiently implemented using the White's wavefunction mapping algorithm,\cite{White1996} where the ground state found during the previous step of the sweep is rotated into the new Hilbert space to be used as a trial state for the diagonalization procedure. This procedure is continued until the shrinking block only contains a single site, and then the direction of the sweep is reversed. Several finite DMRG sweeps are performed until the desired convergence is reached. The finite DMRG algorithm is illustrated in Figure \ref{Figure4}.

By exploiting the sparcity of the block symmetry operators (e.g., the particle-hole and spin-flip symmetries), excited states are conveniently determined by constructing symmetry-adapted states.\cite{doi:10.1063/1.3149536}. Within a symmetry sector higher-lying states are then determined via a Gram-Schmidt projection.
In order to accurately describe these excited states, it is necessary to retain the basis states that optimally represent them  in the truncated Hilbert space. This is achieved by including them in the reduced density matrix, i.e.,
\begin{equation}
	\hat{\rho}_S = \trace_E \sum_i \kappa_i \ket{\Psi_i}\bra{\Psi_i},
\end{equation}
where the summation includes all the targeted states. $\kappa_i$ is usually chosen to be the same for all states such that $\sum_i \kappa_i =1$.

%

\subsection{Adaptive Time-dependent DMRG (tDMRG)}

The dynamics of the evolving system under the Hamiltonian $\hat{H}$ is fully determined by solving the  time-dependent Schr\"odinger equation
\begin{equation}
i \hbar \frac{\partial}{\partial t}\ket{\Psi(t)} = \hat{H}(t)\ket{\Psi(t)}. 	\label{eqn:TDSE}
\end{equation}
In the limit that $\delta t \rightarrow 0$, eq \eqref{eqn:TDSE} has the formal solution
\begin{equation}
	\ket{\Psi(t+\delta t)} = \exp \left (- i\hat{H} (t) \delta t/\hbar \right ) \ket{\Psi(t)}.
\end{equation}

The adaptive time-dependent density matrix renormalization group method, developed in 2004, generalised the DMRG algorithm to study time dependent phenomena.\cite{White2004b,Daley2004a} In this formalism, the evolving state $\ket{\Psi(t)}$ is determined in a truncated Hilbert space such that the loss of information about the system is minimized. The algorithm is efficiently implemented for Hamiltonians  containing only on-site and nearest neighbor interactions. Such a Hamiltonian  can be written as a sum of bond Hamiltonians,
\begin{equation}
\hat{H} = \hat{H}_{1,2} + \hat{H}_{2,3} + \hat{H}_{3,4} +  \cdots \hat{H}_{n,n+1} + \cdots + \hat{H}_{N-1,N},
\end{equation}
where $\hat{H}_{n,n+1}=\frac{1}{2} \left (  \hat{H}_{n}^{\textrm{intra}} + \hat{H}_{n+1}^{\textrm{intra}}  \right ) + \hat{H}_{n,n+1}^{\textrm{inter}}$ acts on the $n^\textrm{th}$ bond. Since neighboring bond Hamiltonians do not commute, a Suzuki-Trotter decomposition is invoked for the propagator, i.e.,
\begin{equation}
	{\rm e}^{-i\hat{H} \delta t/\hbar} \approx {\rm e}^{-i\hat{H}_{1,2} \delta t /2\hbar} {\rm e}^{-i\hat{H}_{2,3} \delta t /2\hbar} \cdots {\rm e}^{-i\hat{H}_{N-1,N} \delta t /2\hbar} {\rm e}^{-i\hat{H}_{N-1,N} \delta t /2\hbar} \cdots {\rm e}^{-i\hat{H}_{2,3} \delta t /2\hbar} {\rm e}^{-i\hat{H}_{1,2} \delta t /2\hbar} + O(\delta t^3)
\end{equation}

The link time evolution operator, $\hat{U}_{n}={\rm e}^{-i\hat{H}_{n,n+1} \delta t /2\hbar}$, is exactly applied on $\ket{\Psi(t)}$ at DMRG step $(n-1)$.\cite{White2004b} At this step the DMRG state  is
 \begin{equation}
\ket{\Psi} = \sum_{l,\alpha_n,\alpha_{n+1},r} \Psi_{l,\alpha_n,\alpha_{n+1},r}
\ket{l_{n-1}} \ket{\alpha_n} \ket{\alpha_{n+1}} \ket{r_{N-(n+1)}}.
\end{equation}
The states $\ket{l_{n-1}}$ and $\ket{r_{N-(n+1)}}$ are eigenvectors of the reduced density matrices corresponding to the system and environment DMRG blocks at step $n$. The states $\ket{\alpha_n}$ and $\ket{\alpha_{n+1}}$ are the exact basis states for sites $n$ and $n+1$.

To find $\ket{\Phi}= \hat{U}_n \ket{\Psi}$, the 2-site augmented block state $\ket{m}=\ket{\alpha_n} \ket{\alpha_{n+1}}$ is transformed to the basis spanned by the eigenstates of $\hat{H}_{n,n+1}$, i.e., $\hat{H}_{n,n+1} \ket{m} \rangle = \epsilon_{m} \ket{m} \rangle$, where
\begin{align}
	\ket{m} \rangle
		&=
			\sum_{m^\prime} \phi_{m^\prime, m} \ket{m^\prime}	.
\end{align}
With this transformation, $\ket{\Phi}= \hat{U}_n \ket{\Psi}$ can be written as,
\begin{equation}
	\ket{\Phi}=
		\sum_{l,m^{\prime \prime},r} \sum_{m, m^\prime}	
		{\rm e}^{-i\epsilon_{m^\prime} \delta t/2\hbar} \phi_{m, m^\prime} \phi_{m^{\prime \prime}, m^\prime}
		\Psi_{l,m,r}
		\ket{l} \ket{m^{\prime\prime}} \rangle \ket{r}.
\end{equation}

\begin{figure}
{
\begin{tikzpicture}[
    font=\sffamily,
    level/.style={black,thick},
    sublevel/.style={black,densely dashed},
    ionization/.style={black,dashed},
    transition/.style={black,->,>=stealth',shorten >=1pt},
    radiative/.style={transition,decorate,decoration={snake,amplitude=1.5}},
    indirectradiative/.style={radiative,densely dashed},
    nonradiative/.style={transition,dashed},
    thick,scale=0.6, every node/.style={scale=0.6},
  ]

%
%
%

  \tikzstyle{every node}=[font=\small]
\coordinate (A1) at (-10,10);
\coordinate (A2) at (-9,9);
\coordinate (B) at (-8,9.5);
\coordinate (C) at (-6,9.5);
\coordinate (D1) at (-5,10);
\coordinate (D2) at (8,9);

  \draw[fill=red] (A1) rectangle (A2);
  \draw (B) circle (0.5cm);
  \draw (C) circle (0.5cm);
  \draw[fill=green] (D1) rectangle (D2);

    \node[align=center,font=\small] at (-9.5,9.5) {$1$};
\node[align=center,font=\small] at (1.5,9.5) {$N-3$};

    \coordinate (P1) at (-9.5,10.2);
   \coordinate (P2) at  (-8,10.2);

  \draw [decorate,
    decoration = {calligraphic brace}] (P1) --  +(1.5,0);
  \draw [decorate,
    decoration = {calligraphic brace}] (P2) --  +(2,0);

  \node[align=center] at (-8.75,11) {$\hat{U}_{1} $};
   \node[align=center] at (-7,11) {$\hat{U}_{2} $};


   \coordinate (A1) at (-10,7);
\coordinate (A2) at (-7,6);
\coordinate (B) at (-6,6.5);
\coordinate (C) at (-4,6.5);
\coordinate (D1) at (-3,7);
\coordinate (D2) at (8,6);

  \draw[fill=red] (A1) rectangle (A2);
  \draw (B) circle (0.5cm);
  \draw (C) circle (0.5cm);
  \draw[fill=green](D1) rectangle (D2);

   \node[align=center,font=\small] at (-8.5,6.5) {$2$};
\node[align=center,font=\small] at (2.5,6.5) {$N-4$};

    \coordinate (P1) at (-7.5,7.2);
   \coordinate (P2) at  (-6,7.2);

  \draw [decorate,
    decoration = {calligraphic brace}] (P2) --  +(2,0);

   \node[align=center] at (-5,8) {$\hat{U}_{3}$};

      \draw [-{Stealth}{Stealth}{Stealth}] (-1,5.9)--(-1,4.5);


  \coordinate (A1) at (-10,3);
\coordinate (A2) at (-3,2);
\coordinate (B) at (-2,2.5);
\coordinate (C) at (0,2.5);
\coordinate (D1) at (1,3);
\coordinate (D2) at (8,2);

  \draw[fill=red] (A1) rectangle (A2);
  \draw (B) circle (0.5cm);
  \draw (C) circle (0.5cm);
  \draw[fill=green] (D1) rectangle (D2);

    \coordinate (P1) at (-3.5,3.2);
   \coordinate (P2) at  (-2,3.2);

  \draw [decorate,
    decoration = {calligraphic brace}] (P2) --  +(2,0);

   \node[align=center] at (-1,4) {$\hat{U}_{n} $};

      \node[align=center,font=\small] at (-6.5,2.5) {$n-1$};
\node[align=center,font=\small] at (4.5,2.5) {$N-n-1$};


  \coordinate (A1) at (-10,0);
\coordinate (A2) at (3,-1);
\coordinate (B) at (4,-0.5);
\coordinate (C) at (6,-0.5);
\coordinate (D1) at (7,0);
\coordinate (D2) at (8,-1);

  \draw[fill=red] (A1) rectangle (A2);
  \draw (B) circle (0.5cm);
  \draw (C) circle (0.5cm);
  \draw[fill=green] (D1) rectangle (D2);

    \coordinate (P1) at (4,0.2);
   \coordinate (P2) at  (6,0.2);

  \draw [decorate,
    decoration = {calligraphic brace}] (P1) --  +(2,0);
  \draw [decorate,
    decoration = {calligraphic brace}] (P2) --  +(1.5,0);

  \node[align=center] at (5,1) {$\hat{U}_{N-2} $};
   \node[align=center] at (6.75,1) {$\hat{U}_{N-1} $};

         \node[align=center,font=\small] at (-3.5,-0.5) {$N-3$};
\node[align=center,font=\small] at (7.5,-0.5) {$1$};

\end{tikzpicture}

}
\caption{Schematic diagram for a single sweep of the adaptive tDMRG algorithm, illustrating the application of the time evolution operator for a time of $\Delta t/2$. The number of sites in each block is shown as text. An open circle represents a single site block. $\hat{U}_{n}={\rm e}^{-i\hat{H}_{n,n+1} \delta t /2\hbar}$.}
\label{Figure5}
\end{figure}
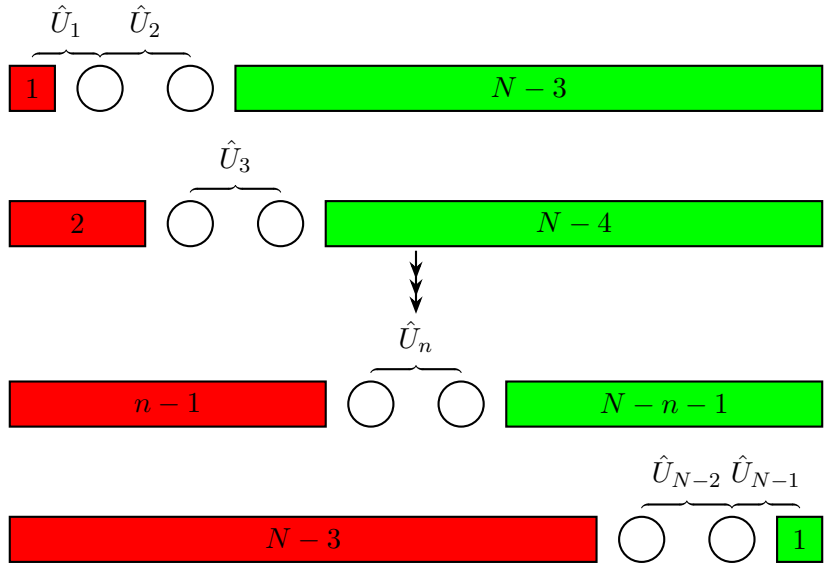

The algorithm now proceeds in precisely the same way as for the static finite lattice DMRG method, namely $\ket{\Phi}$ is truncated via a singular value decomposition and is then transformed to the basis for the next DMRG step via White's wavefunction mapping technique.\cite{White1996} Figure \ref{Figure5} illustrates the key steps of the adaptive tDMRG algorithm.

Readers are referred to refs \citen{Schollwock2005,Schollwock2011a} for a comprehensive review of the DMRG, refs \citen{doi:10.1063/1.5129672,MA202257,MA202291} for recent reviews of applications of DMRG in quantum chemistry, and ref \citen{Schollwock2005a} for a review of time-dependent density matrix renormalization group methods.

\subsection{DMRG accuracy}\label{DMRGacc}

As discussed in section \ref{SDMRG}, the DMRG algorithm finds the optimum $\ket{\tilde{\Psi}}$ by minimizing the quadratic norm $S_2$ (see eq \eqref{eqn:S2}). The truncation error, $\epsilon$, associated with the DMRG algorithm can therefore be defined as the sum of the eigenvalues of $\hat{\rho}_S$ disgarded during the DMRG truncation, i.e.,
\begin{equation}\label{Eq:34}
	\epsilon = \min({S_2}) = 1 - \sum_{\beta = 1}^{M_S} \omega_\beta,
\end{equation}
and $\sum_{\beta = 1}^{r} \omega_\beta = 1$. The reason for the remarkable success of DMRG in explaining the properties of one dimensional quantum systems is understood by the realization that the DMRG truncation error is closely related to the amount of information that is required to accurately represent a quantum system.\cite{Vidal2003c,Vidal2004d} This amount of information is dependent on the entanglement of the system, and is quantified via the von Neumann entanglement entropy,
\begin{equation}\label{Eq:35}
	S
		= - \sum_{\beta=1}^r \lambda_\beta^2 \log \lambda_\beta^2
		= - \Trace_S ( \hat{\rho}_S \log \hat{\rho}_S).
\end{equation}

To illustrate this quantity, first consider a fully unentangled system. This state is described by a product state given by
\begin{equation}
\ket{\Psi} = \left (\sum_{i=1}^{N_S} \Psi_i \ket{i} \right ) \otimes \left (\sum_{j=1}^{N_E} \Psi_i \ket{j} \right ).
\end{equation}
It follows that for this state, $\hat{\rho}_S = \ket{\Phi_S} \bra{\Phi_S}$ where $\ket{\Phi_S}=\sum_{i=1}^{N_S} \Psi_i \ket{i}$, and thus $S=0$. Therefore, the unentangled state has the minimum von Neumann entropy. To demonstrate the maximally entangled state, consider a two spin system described by
\begin{equation}
	\ket{\Psi} = \frac{1}{\sqrt{2}} \left ( \ket{\uparrow_S}\ket{\downarrow_E} + \ket{\downarrow_S}\ket{\uparrow_E} \right ).
\end{equation}
For this state, $\hat{\rho}_S = \frac{1}{2} \left ( \ket{\uparrow_S} \bra{\uparrow_S} + \ket{\downarrow_S} \bra{\downarrow_S} \right )$ and $S=\log 2$. In general, for the maximally entangled state, if the rank of the Schmidt decomposition is $r$, then $\omega_\beta = \lambda_\beta^2 = 1/r$, and $S= \log r$.

Recall from eq \eqref{eqn:Schm} that a general state $\ket{\Psi}$ can be written using a Schmidt basis, where the Schmidt coefficients $\lambda_\beta$ are related to the eigenvalues of $\hat{\rho}_S$ by $\omega_\beta = \lambda_\beta^2$. Thus, if the eigenvalues $\{ \omega_\beta \}$ decay rapidly,  only a small error is introduced by retaining only a relatively small number of eigenstates.

The growth of entanglement with systems size is related to the `area law' which explains why DMRG works so well for one-dimensional systems, but fails in
higher dimensions.\cite{eisert2010colloquium} The area law states that the entanglement entropy of the ground state of a gapped, partitioned system is proportional to the size of the partition between the systems. For one-dimensional systems, the surface between two systems is a point, and as this does not change with system size the entanglement entropy is a constant with system size.\cite{hastings2007area}
This further implies that the truncation error remains essentially constant for a fixed system basis size as a function of system size.
DMRG is thus highly suitable for the study of the electronic states of insulating polyene systems.

\section{Nuclear degrees of freedom}

\subsection{Ehrenfest dynamics}\label{Ehr}

While the electronic wavefunction determined by solving the TDSE is achieved using adaptive tDMRG, the dynamics of the nuclei are determined  via the Ehrenfest equations of motion. The Ehrenfest method assumes that the nuclei evolve on a single effective potential energy surface corresponding to an average of the electronic states contributing to the electronic wavefunction. As a result, despite its mean-field nature, the Ehrenfest method is able to describe transitions between different electronic states.\cite{tully1998}

The nuclear degrees of freedom, defined by eq \eqref{eqn:UVP},  
are treated classically via the Hellmann-Feynman theorem. The force on atom $n$, $f_n$ is

\begin{equation}
f_{\mathrm{n}} = - \frac{\delta \langle \Psi | \hat{H} \ket{\Psi}}{\delta u_{\mathrm{n}}}
	= - \left \langle \frac{\delta \hat{H}}{\delta u_{\mathrm{n}}} \right \rangle.
\end{equation}
This leads to
\begin{equation}
f_{\mathrm{n}} 	= 2 \alpha  \left (	\langle \hat{T}_\mathrm{n}	 \rangle - \langle \hat{T}_{\mathrm{n-1}} \rangle \right	)
	- K  \left  (2 u_{\mathrm{n} } - u_{\mathrm{n} +1} - u_{\mathrm{n} - 1}			\right	). \label{eqn:fn}
\end{equation}
The nuclei obey Newton's equations of motion
\begin{equation}
	\frac{du_{\mathrm{n}}(t)}{dt} = \frac{p_{\mathrm{n}}(t)}{m}
\end{equation}
and
\begin{equation}
	\frac{d p_{\mathrm{n}}(t)}{dt} = f_{\mathrm{n}}(t) - \gamma p_{\mathrm{n}} (t),
\end{equation}
where $p_n$ and $m$ are the nuclear momentum and mass, respectively. A phenomenological linear damping term $\gamma p_{\mathrm{n}}$ is introduced to cause relaxation of the nuclei. The equations of
motion are propagated using the damped Velocity Verlet scheme derived in appendix 5.A of reference \citen{Valentine2020c}, i.e., 
\begin{align}
	u_{n}(t+\Delta t)&=u_{n}(t)+\frac{p_{n}(t)}{m} \Delta t+\frac{1}{2} \frac{[f_{n}(t) - \gamma p_{n}(t)]}{m} \Delta t^{2}	\\
	p_{n}(t+\Delta t)&=\frac{1}{1+\Delta t \gamma / 2}\left[p_{n}(t)[1-\Delta t \gamma / 2]
	+\frac{\Delta t}{2 }\left([f_{n}(t+\Delta t)+f_{n}(t)\right]\right].
\end{align}

%


%

\subsection{Validity of Ehrenfest dynamics} \label{validity}

Molecular dynamics is routinely utilized to simulate numerical dynamics of a diverse range of chemical processes, including gas phase molecular collisions,\cite{Porter1976} reactions in liquids,\cite{allen2017computer} biomolecular dynamics.\cite{Karplus1990} At the heart of the molecular dynamics method is the adiabatic or the Born-Oppenheimer approximation, the assumption that electronic degrees of freedom instantaneously adjust to the nuclear motion such that the electrons move on a single adiabatic surface. Dynamical processes involving electronic transitions violate the Born-Oppenheimer approximation, and as a consequence molecular dynamics is inadequate for the description of dynamical processes involving electronic transitions, and nonadiabatic methods are required.

Ehrenfest dynamics is a non-adiabatic method where the `slow' nuclear degrees of freedom evolve on an effective potential energy surface defined by the weighted average of contributing adiabatic surfaces.\cite{tully1998} The Ehrenfest method allows for feedback between the nuclear and electronic degrees of freedom in both directions, and therefore is a self-consistent method. It is routinely used to model nonadiabatic dynamics.\cite{Micha1982,Sawada1985,Head-Gordon1995,Baa1996,Andrade2009} While Ehrenfest dynamics produce accurate results if the  coupled adiabatic states are similar in character, due to its mean-field nature it fails to properly describe dynamics if the wavepacket bifurcates onto  adiabatic surfaces with very different character.\cite{tully1998}

%

Tully's surface hopping technique was developed to overcome the erroneous mean-field approximation of the Ehrenfest method.\cite{Tully1971} The nuclear degrees of freedom moves on a single potential energy surface until a region of large nonadiabatic coupling is encountered, at which the trajectory splits stochastically. Another approach to improve on the Ehrenfest method is to incorporate quantum effects to the nuclear degrees of freedom.  Mannouch and Barford developed the TEBD-Ehrenfest method for the dynamical simulations of Frenkel-Holstein model.\cite{Mannouch2018,Barford2018c,Mannouch2019} Nuclear motion is described by a mean-field displacement subjected to Ehrenfest dynamics, with quantum fluctuations around the mean-field value. In future work, the TEBD-Ehrenfest algorithm described in ref. \citen{Barford2018c} will be generalized for Debye phonons and its performance will be compared with dynamical simulations based on Ehrenfest equations of motion of refs \citen{Manawadu2022} and \citen{Manawadu2022b}.

\section{Lanczos-DMRG} \label{sec:Lanczos}

Transient (time-resolved) absorption experiments are among the most routinely utilized experimental techniques in the field of carotenoid photophysics.\cite{Berera2009} A typical transient absorption measurement involves photoperturbation of the dynamical system at time $t$ and measurement of the effect of the perturbation at time $t+\tau$. This corresponds to the dynamical correlation function $\tilde{G}_\mu(t,\tau)$ defined as
\begin{equation}
	\tilde{G}_\mu(t,\tau) = \mel{\Psi(t)}{\hat{\mu}(t+\tau) \hat{\mu}(t)}{\Psi(t)},
\end{equation}
where $\hat{\mu}(t)$ is the Heisenberg representation of the dipole moment operator $\hat{\mu}$. Via a Fourier transform of $\omega$ with respect to $\tau$, we obtain the dynamical correlation function in the frequency domain as\cite{jeckelmann2008dynamical}
\begin{equation}
	G_\mu(t,\omega+ i \hbar \eta)
		= - \frac{1}{\pi} \left \langle \Psi(t) \biggl | \hat{\mu} \frac{1}{E_{\Psi} + \omega + i \hbar \eta - \hat{H}}      \hat{\mu} \biggr | {\Psi(t)}  \right \rangle,
\end{equation}
where $E_{\Psi}$ is the energy of $\ket{\Psi (t)}$, and $\eta$ is a small positive real number used to shift the poles of $G_\mu(t,\omega)$ into the complex plane. The imaginary part of $G_\mu(t,\omega+ i \hbar \eta)$ is given by
\begin{equation}
	I_{\mu}(t,\omega + i \hbar \eta)
		= \frac{1}{\pi} \left \langle \Psi(t) \biggl | \hat{\mu} \frac{\eta}{(E_{\Psi} + \omega - \hat{H})^2 + \hbar^2 \eta^2}      \hat{\mu} \biggr | {\Psi(t)}  \right \rangle .
\end{equation}
If a complete set of eigenstates $\{ \ket{n} \}$ of $\hat{H}$, with eigenvalues $\{E_n\}$ is known, using $\hat{1}=\sum_n \ket{n}\bra{n}$, we can write
\begin{equation}
	I_{\mu}(t,\omega + i \hbar \eta)
		= \frac{1}{\pi} \sum_n | \mel{n}{\hat{\mu}}{\Psi(t)}|^2
		 \frac{\eta}{(E_{\Psi} + \omega - E_n)^2 + \hbar^2 \eta^2}     .
\end{equation}
In the limit $\eta \rightarrow 0$, we obtain the transient absorption spectrum at time $t$ via
\begin{equation}
	I_{\mu}(t,\omega)
		= \frac{1}{\pi} \sum_n \left | \mel{n}{\hat{\mu}}{\Psi(t)} \right |^2
		 \delta(E_{\Psi} + \omega - E_n). \label{eq:spectrum}
\end{equation}

It is possible to calculate the transient absorption via a direct evaluation of $\{ \ket{n} \}$ by targeting the $n$ eigenstates in the DMRG density matrix.\cite{Jeckelmann2007a} However, for a fixed number of retained DMRG density matrix eigenstates, the truncation error grows rapidly with the number of targeted states in the density matrix and thus limits the use of this approach. Hallberg,\cite{PhysRevB.52.R9827} and K\"{u}hner and White\cite{Kuhner1999a} introducted the Lanczos-DMRG method which combines DMRG with the Lanczos algorithm.\cite{PhysRevLett.59.2999} Lanczos-DMRG is based on the observation that it is only necessary to calculate the eigenstates that make a finite contribution to the spectrum. This is achieved by projecting the Hamiltonian $\hat{H}$ onto a Krylov subspace spanned by the Lanczos vectors $\ket{f_n}$.

We calculate transient spectra from the state $\ket{\Psi (t)} = \ket{S_i}$ using the Lanczos-DMRG method, with normalised Lanczos vectors defined as in ref \citen{Kuhner1999a}, i.e.,
\begin{align}
	a_n
		&= \mel{f_n}{\hat{H}}{f_n},	\\
	\ket{r_n}
		&= \left( \hat{H} -a_n \right) \ket{f_n} - b_{n-1} \ket{f_{n-1}}, \\
	b_n &= \lVert \ket{r_n} \rVert_2,
\end{align}
and
\begin{equation}
	\ket{f_{n+1}}
		= \frac{\ket{r_n}}{b_n},
\end{equation}
where $b_{-1}=0$. We require $\ket{S_i}$ to be the lowest energy eigenstate of the Hamiltonian in the projected Krylov subspace. This is achieved by projecting out the lower energy eigenstates via
\begin{align}
	\ket{f_0}
		&= \left (\hat{1} - \sum_{j=0}^{i-1} \ket{S_j}\bra{S_j} \right) \ket{g_0},
\end{align}
where
\begin{align}
	\ket{g_0}
		&= \frac{\hat{\mu} \ket{S_i}}{\lVert\hat{\mu} \ket{S_i}\rVert_2}.
\end{align}
Projecting out the eigenstates lower in energy than $\ket{S_i}$ is also motivated as we are only interested in the calculation of stimulated absorption spectra.

We now outline the implementation of Lanczos-DMRG within our adaptive tDMRG simulations. Once the dynamical simulation for time $t$ is reached, we perform a static DMRG sweep to calculate observables, ending at the DMRG step where the system block is the same size as the environment block. We save the full Hilbert space, which is in the tDMRG basis. Then we perform a full static DMRG sweep, diagonalizing the Hamiltonian at each DMRG step to find $\ket{S_i}$, and the Lanczos vectors $\{ \ket{f_n} \}$, as defined above.
The first five Lanczos vectors are used as target states during the static DMRG sweep, weighted proportionately by their contribution to the spectrum as defined by eq (17) of ref \citen{Kuhner1999a}, i.e.,
\begin{equation}
	w_n
		=
		 \sum_m (\Phi_m^0)^2(\Phi_m^n)^2,
\end{equation}
where $w_n$ is the weight of the Lanczos state $n$,  $\Phi_m^n = \braket{f_m}{\Phi_n}$, and $\ket{\Phi_n}$ is the $n^{\textrm{th}}$ eigenstate of the Hamiltonian in the Lanczos Hilbert space. Noting that we need an accurate representation of all eigenstates up to and including $\ket{S_i}$ during the static DMRG sweep, we assign $25\%$ of the weight to $\ket{\Psi(t)}$, $25\%$ of the weight to all eigenstates up to $\ket{S_i}$ (distributed equally), and the remaining $50\%$ of the weight to the target Lanczos states. At the end of the static DMRG sweep, we calculate the transient absorption spectrum using eq \eqref{eq:spectrum}. The dynamical simulation is then continued from the saved tDMRG Hilbert space.

Lanczos-DMRG is a simple and efficient method for the calculation of the low energy discrete absorption spectra, and is well suited for a tDMRG simulation  where the transient absorption is calculated as a function of time during the dynamics. More accurate, but expensive, frequency-space DMRG methods, such as correction vector DMRG\cite{Kuhner1999a} and dynamical DMRG,\cite{Jeckelmann2002a} are required for the calculation of complex spectra with high energy absorptions, or continuous spectra.

\section{Accuracy and convergence}

The adaptive tDMRG algorithm utilized in the calculations of this paper and its companion\cite{Manawadu2022b} suffers from two sources of error. First, the accuracy of the DMRG algorithm is dictated by the truncation error, defined by eq (\ref{Eq:34}). This error is minimized by retaining more states during the truncation. Second, the Suzuki-Trotter decomposition introduces the Trotter error which is reduced by reducing the Trotter time step, $\Delta t$. Minimizing the Trotter time step also minimizes the error associated with the use of Velocity Verlet integrator for the nuclear dynamics.

\subsection{Truncation error and entanglement entropy}

Due to the variational nature of the DMRG algorithm, its accuracy can be  systematically improved by increasing the number of states retained during the DMRG truncation step. Convergence of the closely related PPPP model has been extensively studied.\cite{PhysRevLett.82.1514,PhysRevB.65.075107,PhysRevB.66.205112,doi:10.1063/1.3149536}

	\begin{figure}[h!]

            \centering
         \includegraphics{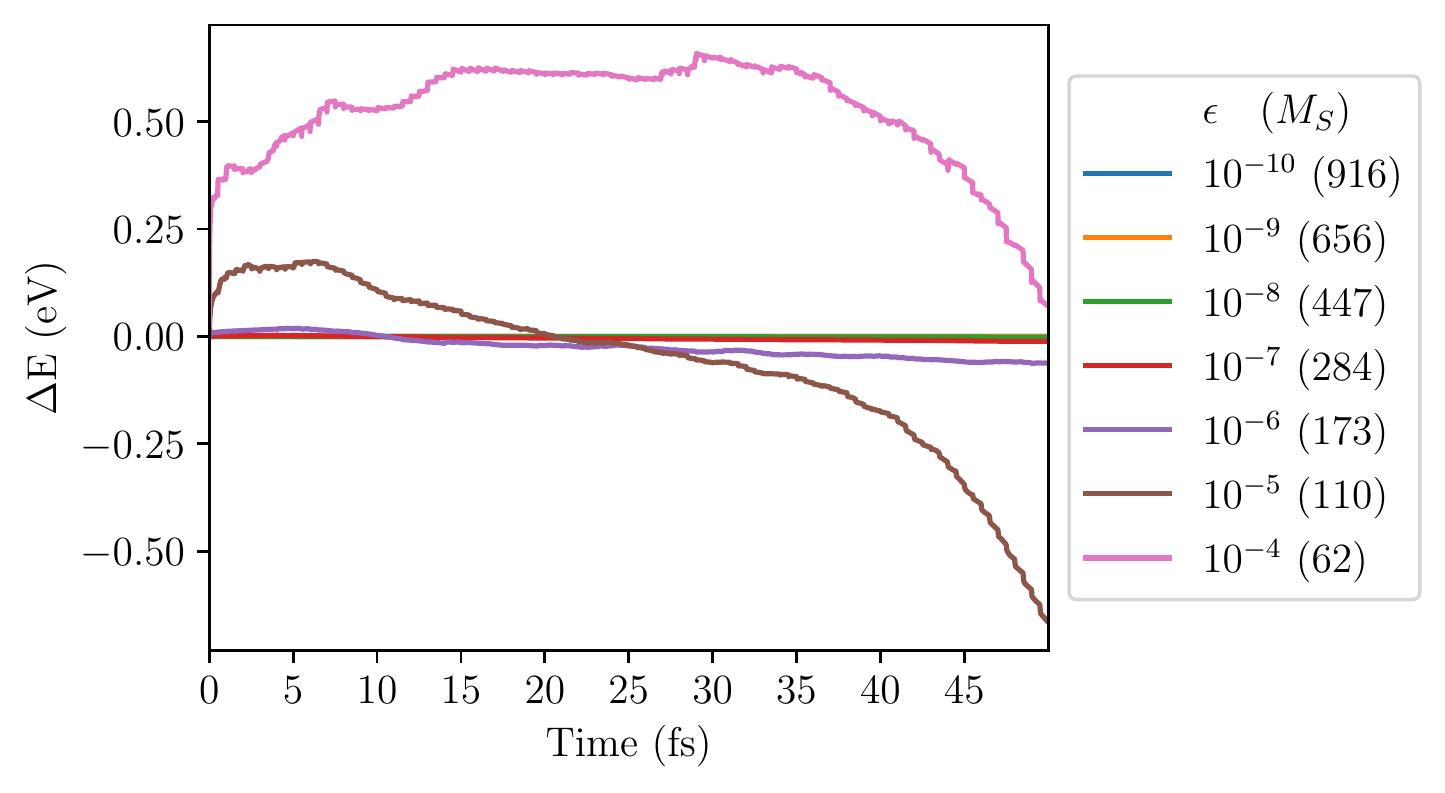}
        \caption{
The calculated difference of the energy as a function of time of the system described by the evolving state vector
$\ket{\Psi(t)}$  for different DMRG truncation errors, $\epsilon$, with respect to the energy calculated with a DMRG truncation error of $10^{-11}$. The maximum number of augmented block states, $M_S$, retained to reach the specified DMRG accuracy is indicated within parenthesis. The evolving state energy is converged for a DMRG truncation error $\epsilon =10^{-8}$.
}
\label{Figure6}
\end{figure}

The convergence of the DMRG truncation error can be determined by evaluating the convergence of an observable as a function of the truncation error. As an illustrative example, Figure \ref{Figure6} shows the variation of energy of the evolving wavefunction $\ket{\Psi (t)}$ with the DMRG truncation error for neurosporene ($N=18$), with $V = 3.25$ eV. The energy is converged for a truncation error of $\sim 10^{-8}$ to an accuracy of $0.1$ eV.

In general, the number of states required to be retained in a DMRG scheme is given by\cite{Feiguin2013}
\begin{equation}
m \approx \exp(S),\label{D3}
\end{equation}
where $S$ is the von Neumann entanglement entropy (defined in eq (\ref{Eq:35})). For our simulations, the maximum von Neumann entanglement entropy reached,
$S_{\textrm{max}}<4.0$, and thus from eq (\ref{D3}) the number of states required to retain $m \approx 55$. The truncation cutoff of $\sim 10^{-8}$ is reached by typically retaining $\sim 400$ augmented block states during the DMRG truncation, which is much larger than the number of states required by eq (\ref{D3}).

\subsection{Trotter error}

\begin{figure}[h!]

            \centering
         \includegraphics{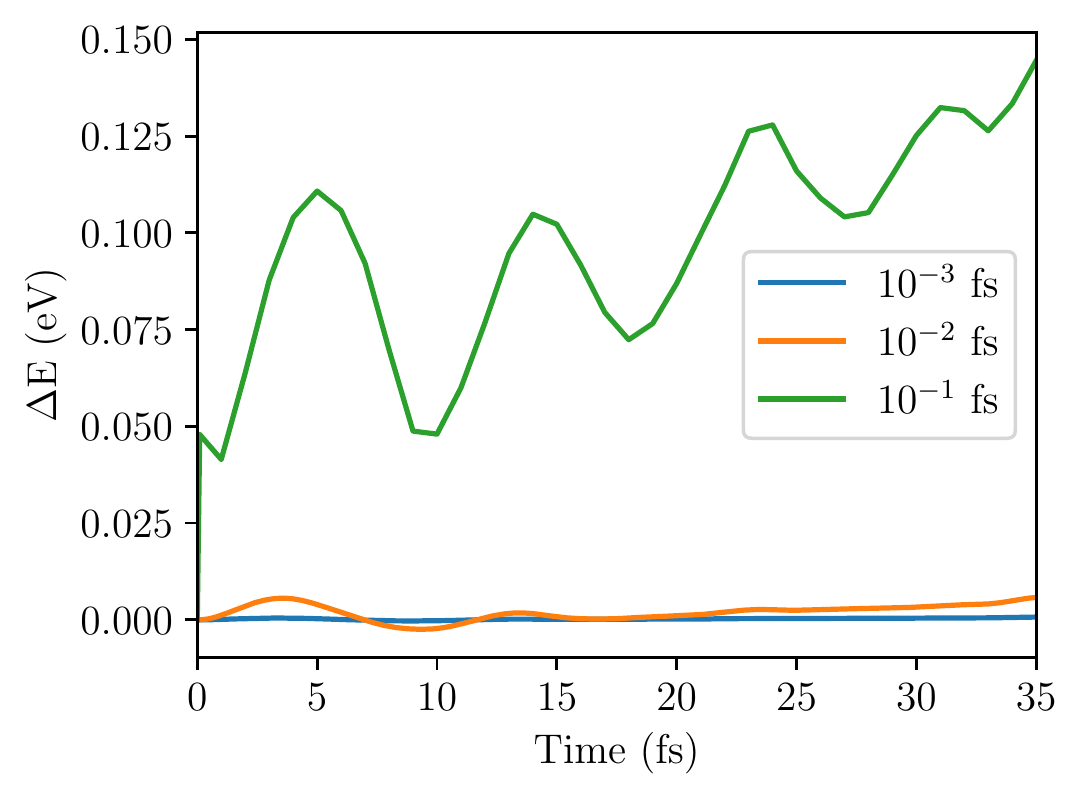}
  \caption{The calculated difference of the energy as a function of time of the system described by the evolving state vector $\ket{\Psi (t)}$  for different Trotter time steps, $\Delta t$, with respect to the energy calculated with a Trotter time step of $10^{-4}$ fs. Convergence is reached with a Trotter time step of $10^{-3}$ fs.}
  \label{Figure7}
  \end{figure}

For an accurate representation of nuclear dynamics, the time step $\Delta t$ should be much smaller than the timescale of nuclear motion, set by carbon-carbon bond oscillations. Typical carbon-carbon bond oscillation frequencies are $\sim 20$ fs, and thus we require $\Delta t <  20 \times 10^{-3}$ fs. Figure \ref{Figure7} illustrates the calculated expectation value of energy of the evolving wavefunction $\ket{\Psi (t)}$ as a function of $\Delta t$, for neurosporene ($N=18$), with $V = 3.25$ eV. We notice that the energy is well converged for $\Delta t =  10^{-3}$ fs, and we use this value as the Trotter time step.

\subsection{Comparison with static DMRG calculations based on Hellmann-Feynman theorem}

\begin{figure}[h!]

            \centering

\includegraphics{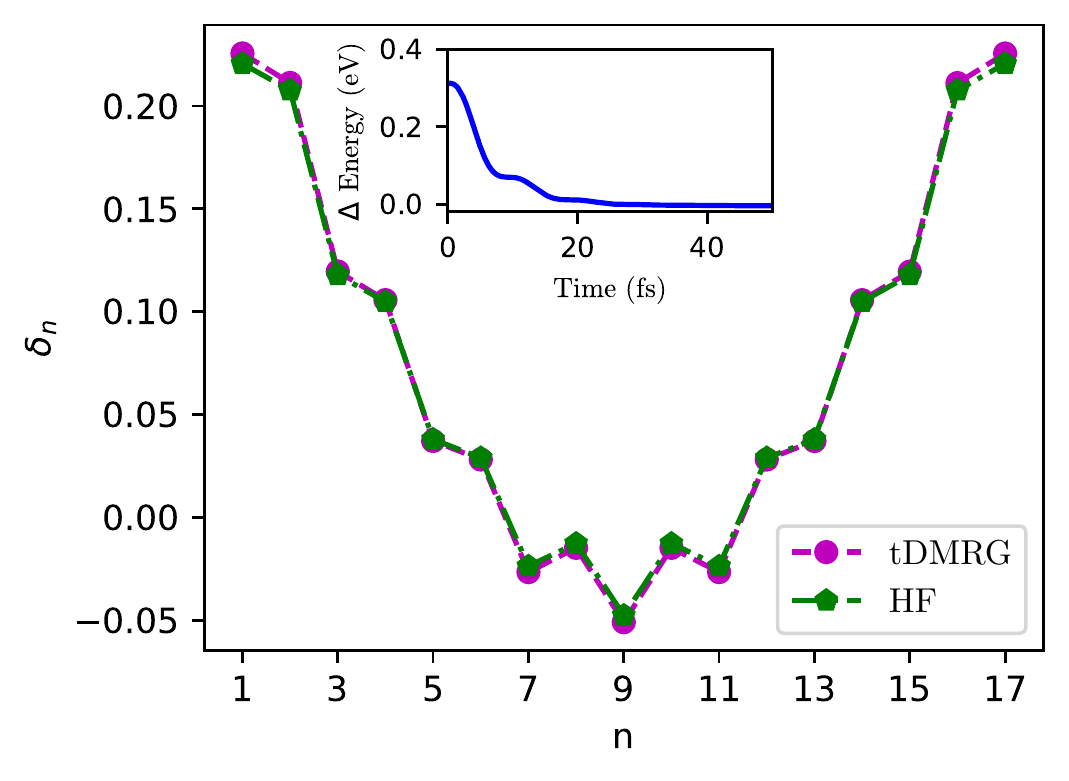}
  \caption{Staggered bond dimerizations of the relaxed $1^1 B_u^+$ state calculated via tDMRG (magenta) and Hellmann-Feynman procedure (green).
  The inset shows the difference between the energy of the $1^1 B_u^+$ state calculated via the adaptive tDMRG and the energy of the $1^1 B_u^+$ state at the relaxed geometry calculated via the Hellmann-Feynman procedure, showing that the two energies converge to the same value.}
  \label{Figure8}
  \end{figure}

Convergence of our program is also evaluated by comparing our dynamical results to static DMRG results obtained via the Hellmann-Feynman theorem. In the absence of the symmetry breaking term, a system prepared in $\ket{\Psi(t=0)}=\ket{1^1 B_u^+}$ at the Franck-Condon point will evolve under linear damping on the $1^1 B_u^+$ potential energy surface, reaching its equilibrium geometry in the long time limit. As the force on atom $n$, $f_n=0$ once equilibrium is reached, the relaxed $1^1 B_u^+$ state can also be found via static DMRG by setting $f_n=0$ in eq \eqref{eqn:fn} and solving the resultant equation iteratively. By comparing observables, we find that the results of our dynamic simulations converge to the same results we obtain via the self iterative Hellmann-Feynman procedure. We illustrate this convergence for the two observables of energy and staggered bond dimerizations.
Figure \ref{Figure8} shows the staggered bond dimerizations, defined for the $n^{\textrm{th}}$ bond by $\delta_n = (-1)^n (\delta u_n - \bar{\delta u_n})/\bar{\delta u_n}$ (where $\delta u_n = u_{n+1} - u_n$ and $\bar{\delta u_n}$ is the average of $\delta u_n$ over all bonds), are converged.
The inset shows that the energy of the $1^1 B_u^+$ state as it evolves reaches the relaxed energy calculated via the Hellmann-Feynman procedure.

\subsection{Accuracy of Lanczos-DMRG calculations}

\begin{figure}[h!]

            \centering
         \includegraphics{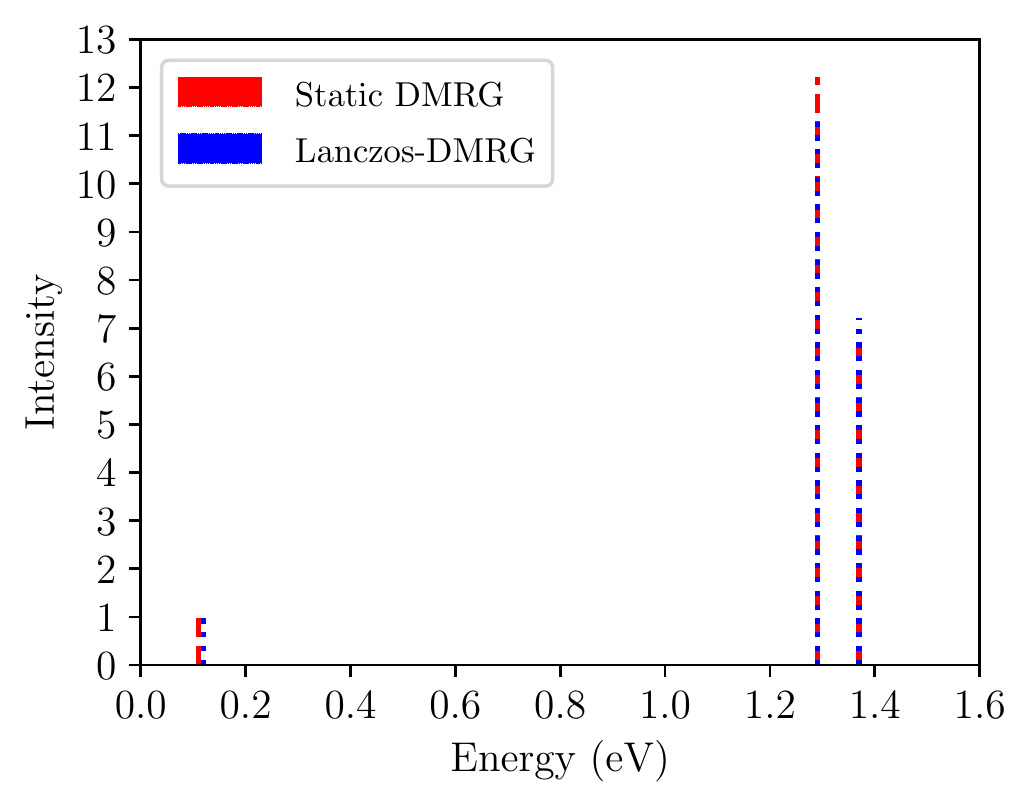}
  \caption{The transient absorption spectra calculated from the $1^1 B_u^+$ state using static DMRG (red) and Lanczos-DMRG (blue). The three signals correspond to transitions to the  $2^1 A_g^-$, $4^1 A_g^-$, and $5^1 A_g^-$ states in  order of increasing energy. The intensity (I) is normalized with respect to the $2^1 A_g^-$ signal, i.e., $I(n^1 A_g^-) = |\mel{n^1 A_g^-}{\hat{\mu}}{1^1 B_u^+}|^2/|\mel{2^1 A_g^-}{\hat{\mu}}{1^1 B_u^+}|^2$. }
  \label{Figure9}
  \end{figure}

The $1^1 B_u^+$ state is connected to the excited states in the $A_g^-$ sector by the dipole moment operator. Therefore, the accuracy of the Lanczos-DMRG can be evaluated by comparing the calculated transient spectra to the energies and transition dipole moments of the excited states in the $A_g^-$ sector, calculated using static DMRG in the absence of symmetry breaking and at the vertical geometry. The results are shown in Figure \ref{Figure9}, and demonstrate a good agreement between the two methods.

Although targeting several Lanczos states during the static DMRG sweep of the Lanczos-DMRG procedure increases the accuracy of the calculation, it leads to a large computational expense. In our simulations, we target five Lanczos vectors, which  maintains the DMRG truncation error at around $\epsilon \sim 10^{-8}$ while keeping $M_S \sim 1000$ augmented block states.

%


\section{Approximate Two Level Dynamics}

\subsection{Quasi-stationary state dynamics}

As observed in ref \citen{Manawadu2022}, for molecules possessing $C_{2}$ symmetry, the optically prepared state $\ket{\Psi (t)}$ is almost entirely composed of two adiabatic states during the entirety of the time evolution. Thus, to a good approximation, we can adopt a two-level system and express $\ket{\Psi(t)}$ as the non-stationary state
\begin{equation}\label{}
  \ket{\Psi(t)} \approx \psi_1 \ket{S_1}\exp(-\textrm{i}E_{1}t/\hbar)  + \psi_2 \ket{S_2}\exp(-\textrm{i}E_{2}t/\hbar),
\end{equation}
where $\ket{S_1}$ and $\ket{S_2}$ are the two contributing adiabatic states, and probability amplitudes $\psi_1$ and $\psi_2$ are assumed to be constant. Denoting the two diabatic states as $\ket{\phi_1}$ and $\ket{\phi_1}$, we can write\footnote{In an exact two-level system, $|a|^2 = |d|^2$ and $|c|^2 = |b|^2$.}
\begin{equation}\label{}
  \ket{S_1} \approx a(t) \ket{\phi_1} + b(t) \ket{\phi_2}
\end{equation}
and
\begin{equation}\label{}
  \ket{S_2} \approx c(t) \ket{\phi_1} + d(t)\ket{\phi_2}.
\end{equation}
Thus, the probability that the system occupies the diabatic state $\ket{\phi_2}$, $P(\Psi (t); \phi_2) = \left | \braket{\phi_2}{\Psi (t)} \right | ^2$, is
\begin{equation}
	P(\Psi (t); \phi_2) =
		| b \psi_1 |^2 + | d \psi_2 |^2 + (b \psi_1)^* (d \psi_2 ) \cos(E_1 - E_2) t/\hbar. \label{eqn:oscillation}
\end{equation}
Eqn.\ (\ref{eqn:oscillation}) describes the observed oscillatory behaviour of the diabatic probabilities of the two-level system (as shown in Figure \ref{Figure10}(b)). Omitting the oscillatory term, we define the `classical' probability of the state $\phi_2$  as
\begin{equation}
	P_{\mathrm{classical}} (\phi_2) =
		| b \psi_1 |^2 + | d \psi_2 |^2  . \label{eqn:classicalP}
\end{equation}

\subsection{Landau-Zener-like transition}\label{Se:7.2}

\begin{figure}[h!]

            \centering
         \includegraphics[height=0.3\textheight]{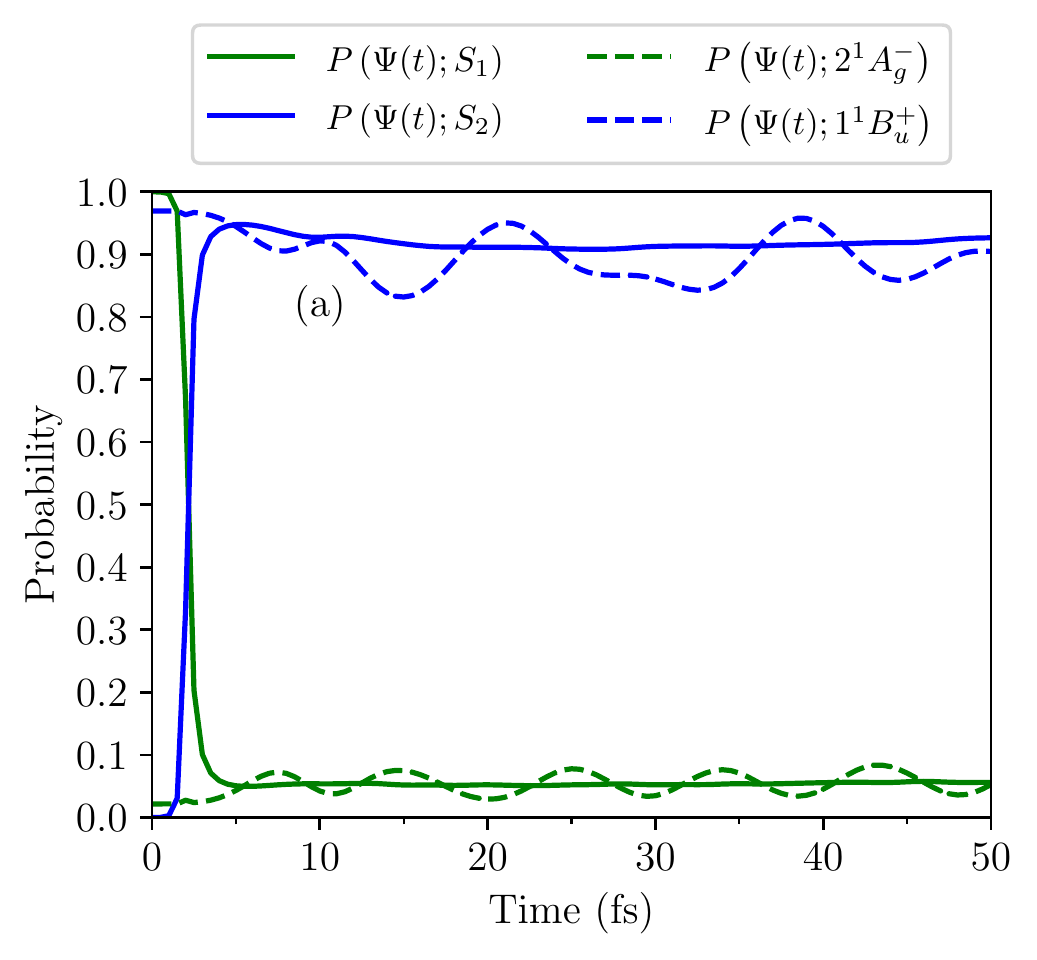}
         \includegraphics[height=0.3\textheight]{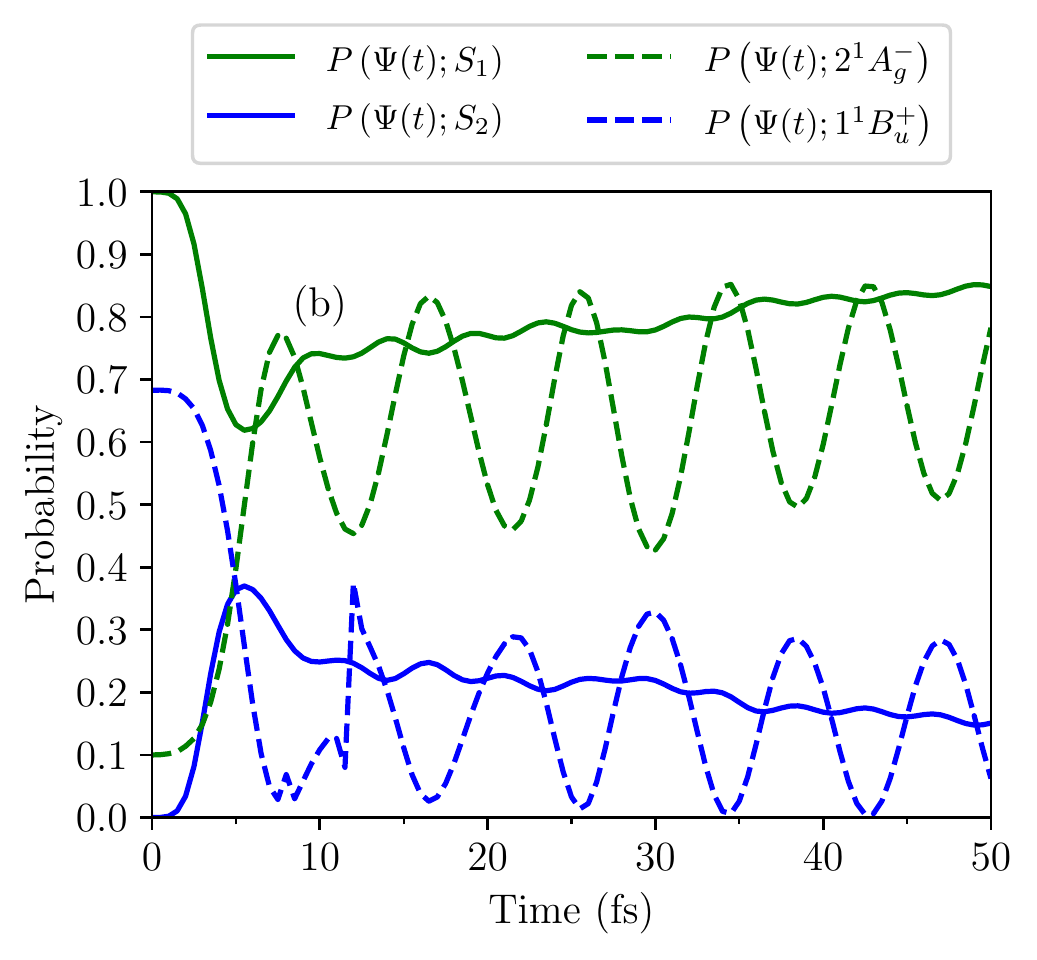}
  \caption{The probabilities as a function of time that the system described by $\ket{\Psi(t)}$ occupies the adiabatic states, $S_1$ and $S_2$ (calculated using eq \eqref{eq:popad}), and the diabatic states, $2^1 A_g^-$ and $1^1 B_u^+$ (calculated using eq \eqref{eq:popdia}).
  The initial condition is that $\ket{\Psi(0)}=\ket{S_1}$.
  Results are for $N=18$ and $V=3.25$ eV, and for the symmetry breaking Hamiltonian presented in Table \ref{Heps} by a linear scaling factor of (a) $\zeta=0.2$ and (b) $\zeta=1.2$.}
  \label{Figure10}
  \end{figure}

In the absence of the particle-hole symmetry breaking term, $\hat{H}_\epsilon$, a system prepared in the $1^1 B_u^+$ state will remain on the same diabatic surface. In this limit, defined by $\epsilon_{\textrm{max}} \rightarrow 0$, the diabatic and adiabatic states are equivalent. Since the diabatic surfaces cross during the dynamical process, it implies that $\ket{\Psi(t)}$ will undergo a nonadiabatic transition while remaining in the $1^1 B_u^+$ diabatic state. However, the inclusion of a non-zero $\hat{H}_\epsilon$ facilitates an adiabatic transition where the system described by $\ket{\Psi(t)}$ transforms from  the $1^1 B_u^+$ diabatic state to either the $1^1 B_u^-$ or $2^1 A_g^-$ states (i.e., negative particle-hole diabatic states) while remaining predominantly in a single adiabatic eigenstate. Figure \ref{Figure10} demonstrates illustrative examples of these two limits. The transition from a `fast' nonadiabatic process to a `slow' adiabatic process as a function of $\epsilon_{\textrm{max}}$ corresponds to a Landau-Zener transition.\cite{zener1932non}
In both cases, although the diabatic surfaces cross, the adiabatic surfaces exhibit an avoided crossing and not a conical intersection.

We study this transition by linearly scaling the potential energies in the symmetry breaking Hamiltonian (presented in Table \ref{Heps}) by a scaling factor $\zeta$. For this system, the $\ket{1^1 B_u^+} \rightarrow \ket{2^1 A_g^-}$ transition is expected energetically (see Figure \ref{Figure2} for the diabatic vertical and relaxed excited energies). The system is prepared in $\ket{\Psi(t=0)} = \ket{S_1}$, the dipole-allowed adiabatic excited state.

	\begin{figure}[h!]

            \centering
         \includegraphics{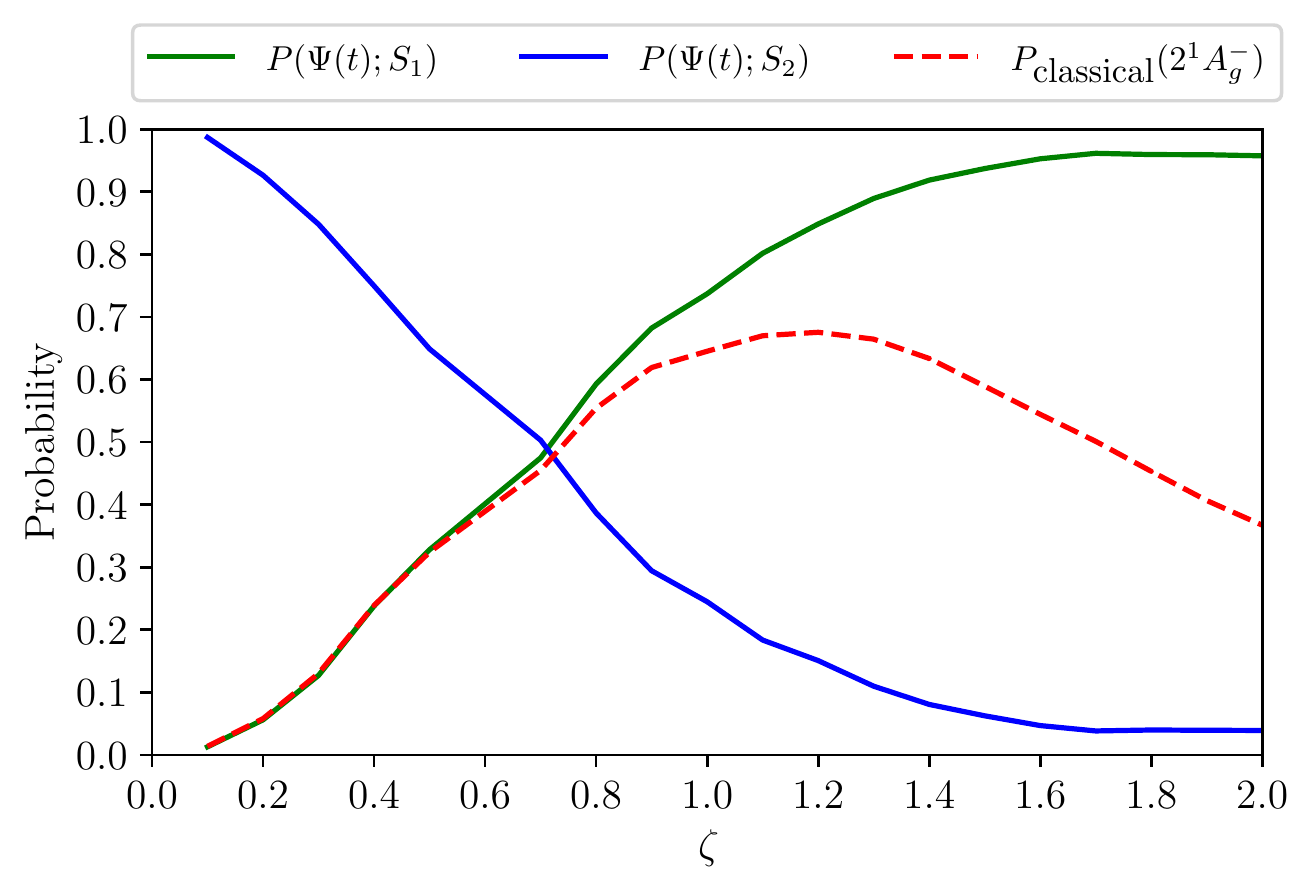}
  \caption{The probabilities at $t=50$ fs as a function of increasing the strength of the particle-hole symmetry term (i.e., $\zeta$) that the system described by $\ket{\Psi(t)}$ occupies the adiabatic states, $S_1$ and $S_2$.
  Also shown is the classical $2^1 A_g^-$ yield.
  The initial condition is that $\ket{\Psi(0)}=\ket{S_1}$. Results are for $N=18$ and $V=3.25$ eV.}
  \label{Figure11}
  \end{figure}

Figure \ref{Figure11} illustrates the probabilities that the system described by $\ket{\Psi(t)}$ occupies the adiabatic excited states $S_1$ and  $S_2$ at $t = 50$ fs as a function of $\zeta$, given the initial condition that $\ket{\Psi(t=0)}= \ket{S_1}$. As $\zeta \rightarrow 0$, $P(\Psi(t \rightarrow \infty); S_2) \rightarrow 1$ and $\ket{S_2}$ is entirely composed of $\ket{1^1 B_u^+}$, as seen from Figure \ref{Figure10}(a). Since the system is prepared with $P(\Psi(t=0); S_1)=1$ this corresponds to a nonadiabatic transition while staying on the $1^1 B_u^+$ diabatic potential energy surface.  As $\zeta$ increases, $P(\Psi(t \rightarrow \infty); S_1)$ increases and as $\zeta \rightarrow 2$, $P(\Psi(t \rightarrow \infty); S_1) \rightarrow 1$, i.e., $\ket{\Psi(t)}$ stays predominantly in $\ket{S_1}$ during the dynamics, while the probability that $\ket{\Psi}$ remains in $\ket{1^1 B_u^+}$ decreases, as seen in Figure \ref{Figure10}(b). Now the transition resembles an adiabatic process.

The `classical' probability, defined by eq (\ref{eqn:classicalP}), i.e.,
\begin{align}
	P_{\mathrm{classical}} (2^1 A_g^-) =& P(S_1;2^1 A_g^-) \times P( \Psi(t);S_1) +
					P(S_2;2^1 A_g^-) \times P( \Psi(t);S_2),
\end{align}
provides a measure of the adiabaticity of the transition. We see that as the particle-hole symmetry breaking term increases (i.e., as $\zeta$ increases), $P_{\mathrm{classical}}$ gradually increases and reaches a maximum $\sim 65 \%$ around $\zeta = 1.2$. As $\zeta$ further increases, $P(\Psi(t); 1^1 B_u^+)+P(\Psi(t); 2^1 A_g^-) < 1$, i.e., $\ket{S_1}$ gains contributions from higher energy diabatic states. Consequently, the process can no longer be modelled as a two-level system.

\section{Conclusions}

This paper has  presented a dynamical simulation scheme to model the highly correlated excited state dynamics of linear polyenes. It complements our more explanatory discussion of carotenoid excited state dynamics in refs \citenum{Manawadu2022} and \citenum{Manawadu2022b}.
We applied it to investigate the internal conversion processes of carotenoids following  their photoexcitation. We use the extended Hubbard-Peierls model, $\hat{H}_{\textrm{UVP}}$, to describe the $\pi$-electronic system coupled to nuclear degrees of freedom. This is supplemented by a Hamiltonian, $\hat{H}_{\epsilon}$, that explicitly breaks both the particle-hole and two-fold rotation symmetries of idealized carotenoid structures. The electronic degrees of freedom are treated quantum mechanically by solving the time-dependent Schr\"odinger equation using the adaptive tDMRG method,  while nuclear dynamics are treated via the Ehrenfest equations of motion. By defining adiabatic excited states as the eigenstates of the full Hamiltonian $\hat{H}=\hat{H}_{\textrm{UVP}}+\hat{H}_{\epsilon}$, and diabatic excited states as eigenstates of $\hat{H}_{\textrm{UVP}}$, we present a computational framework to monitor the internal conversion process from the initial photoexcited $1^1 B_u^+$ state to the singlet triplet-pair states of carotenoids.
We further incorporate Lanczos-DMRG to the  tDMRG-Ehrenfest method to calculate transient absorption spectra from the evolving photoexcited state.  We describe in detail the accuracy and convergence criteria for DMRG, and show that this method  accurately describes the dynamical processes of  carotenoid excited states. We also discuss the effect of the symmetry breaking term, $\hat{H}_{\epsilon}$, on the internal conversion process, and show that its effect  on the extent of internal conversion can be described by a Landau-Zener-type transition.

Mindful of the possible failures of the Ehrenfest approximation at avoided crossings (or conical intersections) described in section \ref{validity}, future work will use the adaptive tDMRG method to simulate the excited state dynamics with fully quantized phonon degrees of freedom.

\newpage

\begin{acknowledgement}

We thank Max Marcus for helpful discussions. D.M is grateful to the EPSRC Centre for Doctoral Training, Theory and Modelling in Chemical Sciences, under Grant No. EP/L015722/1 and Linacre College for a Carolyn and Franco Gianturco Scholarship and the Department of Chemistry, University of Oxford for financial support. D.J.V. received financial support from the EPSRC Centre for Doctoral Training, Theory and Modelling in Chemical Sciences (Grant ref. EP/L015722/1), the Department of Chemistry, and Balliol College Oxford via the Foley-Béjar Scholarship. We acknowledge the use of University of Oxford Advanced Research Computing (ARC) facility for this work.

%

\end{acknowledgement}


%




\newpage

\bibliography{D.Phil-Writing-Theory_semi_review-References}

\end{document}